\documentclass[twocolumn]{aastex631}

\defcitealias{Guilera22a}{Guilera et al. 2022}
\usepackage{amsmath}

\usepackage{graphicx,color}

\begin{document}

\title{Quantifying the Impact of the Dust Torque on the Migration of Low-mass Planets II: \\
The Role of Pebble Accretion in Planet Growth within a Global Planet Formation Model}

\correspondingauthor{Octavio M. Guilera}
\email{oguilera@fcaglp.unlp.edu.ar}

\author[0000-0001-8577-9532]{Octavio M. Guilera}
\affiliation{Instituto de Astrof\'{\i}sica de La Plata (IALP), CCT La Plata-CONICET-UNLP, Paseo del Bosque S/N, La Plata,Argentina.}

\author[0000-0002-3728-3329
]{Pablo Benitez-Llambay}
\affiliation{Facultad de Ingenier\'ia y Ciencias, Universidad Adolfo Ib\'a\~nez, Av. Diagonal las Torres 2640, Pe\~nalol\'en, Chile.}

\author[0000-0001-8031-1957
]{Marcelo M. Miller Bertolami}
\affiliation{Instituto de Astrof\'{\i}sica de La Plata (IALP), CCT La Plata-CONICET-UNLP, Paseo del Bosque S/N, La Plata,Argentina.}

\author[0000-0001-8716-3563
]{Martin E. Pessah}
\affiliation{Niels Bohr International Academy, Niels Bohr Institute, Blegdamsvej 17, DK-2100 Copenhagen Ø, Denmark}

\begin{abstract}
Although dust constitutes only about 1\% of the mass in a protoplanetary disk, recent studies reveal its substantial impact on the torques experienced by low- and intermediate-mass planetary cores. In this study, we present the first comprehensive analysis of the dust torque's influence on the evolution of growing planetary embryos as they migrate through a protoplanetary disk and undergo gas and pebble accretion. Our global model incorporates viscous accretion and X-ray photoevaporation effects on the gaseous disk while also accounting for the dynamic processes of dust growth and evolution, including coagulation, drift, and fragmentation. Our findings demonstrate that dust torque significantly affects planetary migration patterns, particularly facilitating prominent outward migration for planets forming within the water ice line. This outward thrust arises from an enhanced dust-to-gas mass ratio in the inner disk, driven by the inward drift of pebbles from the outer regions. Conversely, for planets that originate beyond the water ice line, while the dust torque attenuates inward migration, it does not substantially alter their overall migration trajectories. This is attributed to the rapid reduction in dust-to-gas mass ratio, resulting from swift pebble drift and the short formation timescales prevalent in that region. Overall, our findings highlight the critical role of dust torque in shaping the migration of low- and intermediate-mass planets, particularly in conditions where increased dust concentrations amplify its effects. These insights have significant implications for understanding the formation timescales, mass distributions, and compositional characteristics of emerging planetary systems.
\end{abstract}


\section{Introduction} 
\label{sec:intro}

Planet formation is a complex phenomenon that involves multiple physical processes occurring simultaneously, each playing a crucial role \citep[see, e.g.,][for recent reviews about planet formation]{Venturini20Review, Drazkowska2023PPVII, MordasiniBurn2024}. Several of these processes are closely interconnected. Notably, the growth of low- and intermediate-mass planets is heavily influenced by the evolution of both the gaseous and solid components in the protoplanetary disk, particularly within the framework of pebble accretion \citep[e.g.,][]{Lambrechts19, Liu2019, Ogihara2020, Venturini20ST, Venturini20Letter, Drazkowska21}. The accretion of gas and solids is significantly affected by the orbital evolution of the growing planets, which is governed by their dynamical interactions with the disk in which they are embedded
\citep[e.g.,][]{Paardekooper2011, Lega2014,  Benitez-llambay2015, jm2017}. 

Most planet formation models including planet migration consider prescriptions that have been designed to  account for the interactions between a planet and a gaseous disk \citep[e.g.,][]{IdaLin04, Alibert05, Dittkrist2014,Guilera2017b, Bitsch19, Izidoro2021, Voelkel2022}. Only a handful of models include the effects of thermal-disk diffusivity and heat released due to solid accretion onto the planet, both of which have an impact on planet migration in a gaseous disk \citep[e.g.,][]{Guilera2019, Venturini20Letter, Baumann2020, Guilera2021, Cornejo+2023, Cornejo+2023b, ChrenkoChametla2023}. 

Even though the solid component is critical in most processes involved in planet formation and evolution, its role in planet migration has only been highlighted recently by \citet[][hereafter BLP18]{BL2018}, who showed that the solid component can generate a non-negligible torque even when contributing to the total disk mass at the standard 1\% level 
\citep{Drazkowska2023PPVII}. This so-called ``dust torque'' is generally positive over a wide range of Stokes numbers and planet masses and has the potential to delay, and even reverse, the planet's inward migration. This basic conclusion has also been confirmed by more recent works, which computed torque maps including the dynamical effects of pebble accretion onto the planet \citep{Regaly2020,Chrenko+2024} and also through linear analysis \citep{Hou2024, Hou2025}. 

In a recent paper, we developed a framework to integrate dust torque into formation track models, utilizing torque measurements from BLP18 \citep{Guilera+2023}. We calculated the migration tracks of individual planets as they evolved embedded in various disk models based on the standard alpha-disk model. To gain insight into how different physical parameters influence the results, we examined the effects of independently varying the particles' Stokes number, the dust-to-gas mass ratio, the stellar mass-accretion rate, and the disk viscosity. Through a series of simulations, employing simple models for the evolution of the disk and the mass accretion onto the planet, we found that the dust torque can significantly impact the evolution of planetary embryos, leading to outward planet migration in many cases. 

The current paradigm for planet formation relies heavily on planetesimals embedded in dusty disks gaining mass through pebble accretion
\citep[see][for a recent review]{Ormel2024}. This scenario involves combining elements that can lead to the impinging dusty flow to develop an asymmetric distribution around these accreting bodies, exerting a net torque on them (BLP18).
In other words, planetesimals growing via pebble accretion are naturally subject to torques resulting from the flow of dust and pebbles passing by and accreting onto them. Currently, however, there are no global models of planet formation --based on pebble accretion-- which include the orbital evolution of the growing planets as determined by the interaction between the planet, the gas, and the solid component in the protoplanetary disk.

Motivated by the need for a more consistent description of planet formation and evolution as driven by pebble accretion, in this work we incorporate the dust torque component introduced in BLP18 into PLANETALP, a global model for planet formation developed by the Planetary Astrophysics Group at La Plata \citep{Ronco2017, Guilera2017b, Guilera2019, Venturini20ST, Guilera2021}. This enables us to study and quantify the importance of the dust torque on the migration of low- and intermediate-mass planets growing by pebble accretion employing more realistic models for disk evolution. In brief, PLANETALP computes the global evolution of the gaseous disk as driven by viscous accretion and X-ray photoevaporation, and integrates the orbital evolution of an embedded planet as it grows in mass due to pebble and gas accretion including a detailed model for dust growth and evolution. The primary aim of this paper is to examine the implications of incorporating the effect of the dust torque into these state-of-the-art global disk models for the first time.

The paper is organized as follows: 
in Sec.~\ref{sec_2}, we briefly describe our global model; 
in Sec.~\ref{sec_3}, we compute planet formation tracks using the dust torque models developed by 
\cite{Guilera+2023}, as based on the results of BLP18 which did not include the dynamical effect of pebble accretion in their torque maps; the impact of this effect is investigated 
in Sec.~\ref{sec_4} by employing the recent torque maps computed by \cite{Chrenko+2024}; in Sec.~\ref{sec_5}, we summarize the main outcomes and discuss their implications. We conclude in Sec.~\ref{sec_6} by highlighting the key takeaway from this work.

\section{Brief Description of the Model}
\label{sec_2}

PLANETALP computes the evolution of a 1D+1D axisymmetric gaseous disk by viscous accretion and X-ray photoevaporation due to the central star. The disk is considered in vertical hydrostatic equilibrium. We solve the structure and transport equations, considering the viscous dissipation and the irradiation from the central star as heating sources. This allows us to evolve the thermodynamic structure of the disk at the mid-plane from which we compute the disk temporal evolution \citep[see][for details]{Guilera2017b, Guilera2019}. 

We highlight here some of the key considerations defining our approach:

{\it Evolution of Solids} \/--
The evolution of the dust and pebbles is computed following \citet{Guilera20}, using a discrete size distribution with 200 size-bins between 1~$\mu$m and a maximum size. At each radial bin, the dust grows to a maximum size that is limited by coagulation, radial drift, and fragmentation --when viscosity becomes very low, fragmentation can be driven by differential drift  \citep{Birnstiel12}. We use the particle mass-weighted mean drift velocities and the mass-weighted mean Stokes numbers to compute the time evolution of the solid surface density solving an advection-diffusion equation \citep[see, e.g.,][]{Drazkowska16}. 

\begin{figure}
    \includegraphics[width=\columnwidth]{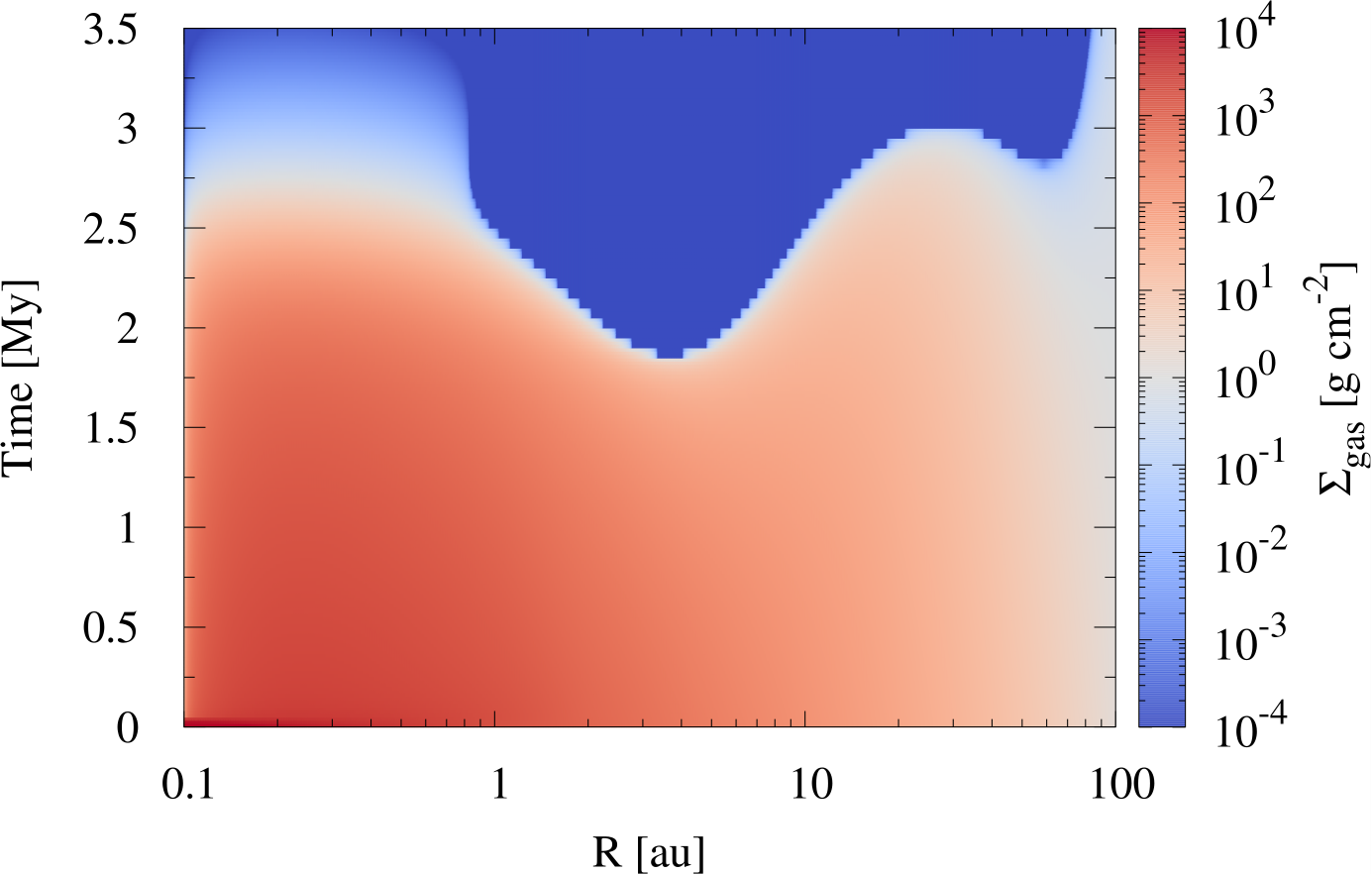} 
    \includegraphics[width=\columnwidth]{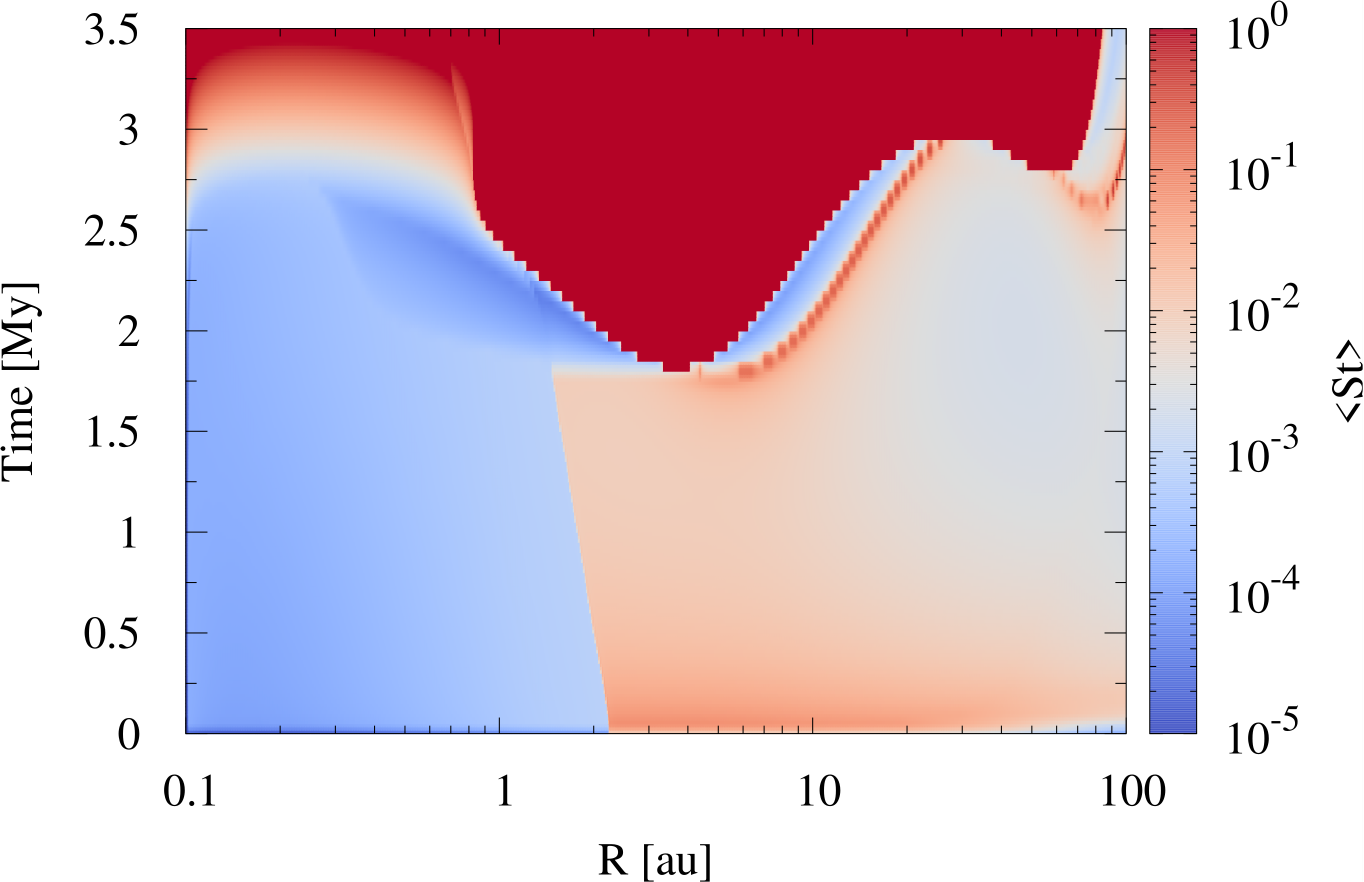} \\
    \includegraphics[width=\columnwidth]{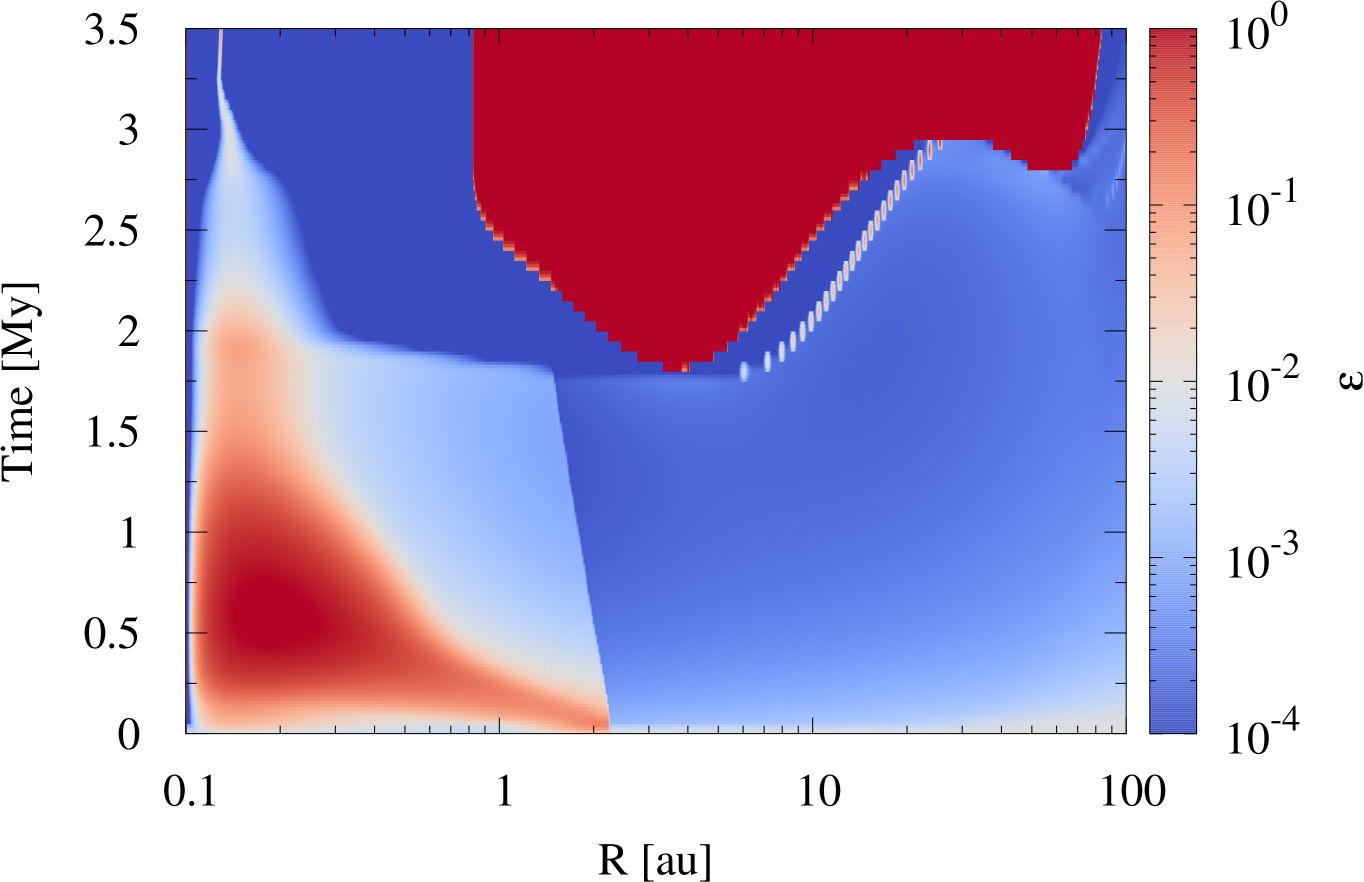}
    \caption{Time evolution maps of the gas surface density (top), particle mass-weighted mean Stokes numbers (middle), and dust-to-gas mass ratio, $\epsilon= \Sigma_{\text{dust}}/\Sigma_{\text{dust}}$ (bottom).} 
    \label{fig1_sec3.1}
\end{figure}

{\it The Water Ice-line} \/--
Our model considers that the dust properties change at the water ice-line\footnote{The water ice-line is defined as the location in the disk where the mid-plane temperature equals 170 K. This location moves inwards as the disk evolves and cools down, see Figure~\ref{fig1_sec3.1}}. We assume that silicate particles have a threshold fragmentation velocity of 1~m/s inside the water ice-line, while the ice-rich particles beyond the water ice-line have a threshold fragmentation velocity of 10~m/s \citep{Gundlach2015, MusiolikWurm2019, Musiolik2021}. The accreted material by the planets and the ice sublimation when the dust/pebbles cross the water ice-line are considered as sink terms, reducing the dust/pebble surface density \citep[see][for details]{Guilera20, Venturini20ST}. 

{\it Accretion of Solids} \/--
As in \citet{Venturini20ST} and \citet{Guilera2021}, we consider that planets grow by the concurrent accretion of surrounding pebbles and gas. Pebble accretion is computed in the 2D and 3D regime considering the pebble scale height \citep{Lambrechts14, Brasser17}. The accretion of pebbles is halted when the planets reach the pebble isolation mass \citep{Lambrechts14, Bitsch18, Ataiee18}. Gas accretion is considered in both the attached and detached phases, following the approach of \citet{Venturini20ST}. During the attached phase, when the planet’s envelope remains smoothly connected to the surrounding gaseous disk, the accretion process is modeled by solving the one-dimensional, spherically symmetric structure equations for the envelope, up until the planet reaches the pebble isolation mass. At this point, solid accretion ceases, and the core becomes critical. In this regime, where solid accretion is no longer occurring, we employ the prescription from \citet{Ikoma00}. As the planet surpasses the pebble isolation mass, its gas accretion rate may exceed the supply regulated by the disk’s viscosity. This leads to the detachment of the planet from the disk, while accreting gas at a rate dictated by viscosity, i.e., $3\pi\nu\Sigma_{\rm gas}$. Additionally, in the detached phase, gas accretion can be further suppressed if the planet opens a gap in the disk \citep{Tanigawa07}. As in \citet{Guilera+2023}, we adopt the conservative approach of allowing planets to grow until they reach $10~M_{\oplus}$ since, as shown in BLP18, the dust torque becomes inefficient beyond this mass range, when the planet opens a gap in the dust.

{\it Torques in a Dusty Disk} \/--
The total torque exerted on a planet embedded in a gaseous disk with a solid component is given by
\begin{eqnarray}
\Gamma_{\text{tot}}= \Gamma_{\text{g}} + \Gamma_{\text{d}}.
\label{eq3_sec2}
\end{eqnarray}
Here, $\Gamma_{\text{g}}$ is the sum of the Lindblad and corotation torques exerted by the gas, for which we adopt the prescriptions given by \citet{jm2017}. For the dust torque, denoted as $\Gamma_{\text{d}}$, we utilize the global models created by \cite{Guilera+2023}. They generated smooth torque maps through a two-dimensional linear interpolation based on the planet's mass and the mass-weighted mean Stokes number\footnote{In Appendix~\ref{appex1}, we compare the results of the model using the dust distribution instead the mass-weighted mean Stokes number to compute the dust torque.} at the planet's location, using the torque maps calculated by BLP18 for a reference disk radius. The detailed computations and numerical implementations of the migration of planets in PLANETALP can be found in \citet{Guilera2019}. 

{\it Dust-to-gas Mass Ratio} \/--
As remarked by BLP18, because the dust torque scales with the dust surface density, the net dust torque in our model is computed multiplying the original dust torque from the bilinear interpolation of the table by a factor $\epsilon_{\text{P}}/\epsilon_0$, where $\epsilon_{\text{P}}$ is the dust-to-gas mass ratio at the planet location and $\epsilon_0= 0.01$. To avoid extrapolating outside the grid where the dust torques were originally computed in BLP18, if the planet mass and the mass-weighted mean Stokes numbers at the planet location are lower than the minimum values of the grid, we adopt these minimum values to compute the dust torque. As explained below, we also note that the mass-weighted mean Stokes numbers at all radii are always lower than unity throughout the disk evolution.

\section{Planet Formation Tracks}
\label{sec_3}

We adopt as our fiducial case study the disk model employed by \citet{Guilera2021}. Our disk has an initial gas surface density given by \citep{Andrews10}
\begin{equation}
\Sigma_{\text{gas}} = \Sigma_{\text{gas}}^0 \left( \dfrac{R}{R_c} \right)\,e^{-(R/R_c)}, 
\end{equation}
$\Sigma_{\text{gas}}^0$ being a normalization constant that depends on the disk mass, and, $R_c= 39$~au, is the characteristic radius of the disk. We also consider a viscosity parameter $\alpha = 10^{-4}$ with an initial mass of $M_{\rm disk} = 0.05~M_{\odot}$ around a central solar-mass star, i.e., $M_\star = M_{\odot}$ with an initially constant dust-to-gas mass ratio $\epsilon (t=0~\text{yr}) = 0.01$.
To understand the impact of the total disk mass and its metallicity, we consider in addition two complementary disk models: {\it i}) the fiducial case with a more massive disk, $M_{\rm disk}=0.1 M_{\odot}$ and {\it ii}) the fiducial case with an initial metallicity $\epsilon (t= 0~\text{yr}) = 0.02$ \footnote{In all cases we check the gravitional stablity of the disk by computing the Toomre-paremeter \citep{Toomre1964} at every timestep and every radial bin \cite[see][for details]{Ronco2017}. All disks presented in this work are gravitationally stable.}. In what follows, we compute planet formation tracks for individual planets starting from an initial core of one Moon mass and with a negligible envelope, initially located at different disk radii. We usually consider 
$r_{\rm p}(t= 0~\text{yr}) = \{0.5, 1, 2, 3, 5, 8, 10, 12\}$ au.

\subsection{Evolution of Fiducial Disk Model}
\label{sec_3.1}

Since the dust torque acting on the planet depends on the pebble Stokes number and the dust-to-gas mass ratio at the planet location, it is necessary to compute in detail the evolution of the solid component throughout the disk.  In Figure~\ref{fig1_sec3.1}, we plot the time evolution maps of the gas surface density (top panel), the particle mass-weighted mean Stokes numbers of the dust distribution (middle panel), and the dust-to-gas mass ratio (bottom panel) for the fiducial disk. The results shown correspond to a  simulation considering the evolution of a disk without a planet embedded in it. From the gas surface density map, it is evident that X-ray photoevaporation opens up a gap in the gas disk at about 2~Myr at a few au, and that the inner part of the disk is depleted of material in about 1~Myr after the gap opening.  

The most important feature to highlight is the rapid change in the dust-to-gas mass ratio on either side of the water ice-line (initially situated at approximately 2.25 au). This variation is primarily driven by the drift of particles from the outer regions of the disk toward the inner regions. Within the first million years of disk evolution, the dust-to-gas mass ratio experiences an increase of more than an order of magnitude inside about 1 au, while it experiences a notable decrease beyond the water ice line.

As the threshold fragmentation velocity of the ice-rich particles is larger beyond the water ice-line, they reach larger sizes and Stokes numbers with respect to the silicate particles inside the water-ice line. This leads to mass-weighted mean Stokes numbers that are larger beyond the water ice-line. These Stokes numbers reach values between approximately $10^{-2}$ and $10^{-1}$ until the gap in the gas disk is formed. In contrast, within the water ice line (clearly distinguishable as the boundary between the red and blue color scales in Figure~\ref{fig1_sec3.1}), particles grow to smaller sizes, resulting in mass-weighted mean Stokes numbers in the range of $10^{-4}$ to $10^{-3}$. After the gap forms, there is a marked decrease in the surface densities of gas and dust/pebbles, particularly in the inner region of the disk where material supply is limited. Consequently, dust and pebbles accumulate primarily in the pressure maximum created by the internal photoevaporation at the edge of the outer disk where the aerodynamical drag force vanishes.

\subsection{Planet Formation Tracks in Fiducial Disk Model}
\label{sec_3.2}

\begin{figure}
    \centering
    \includegraphics[width=\columnwidth]{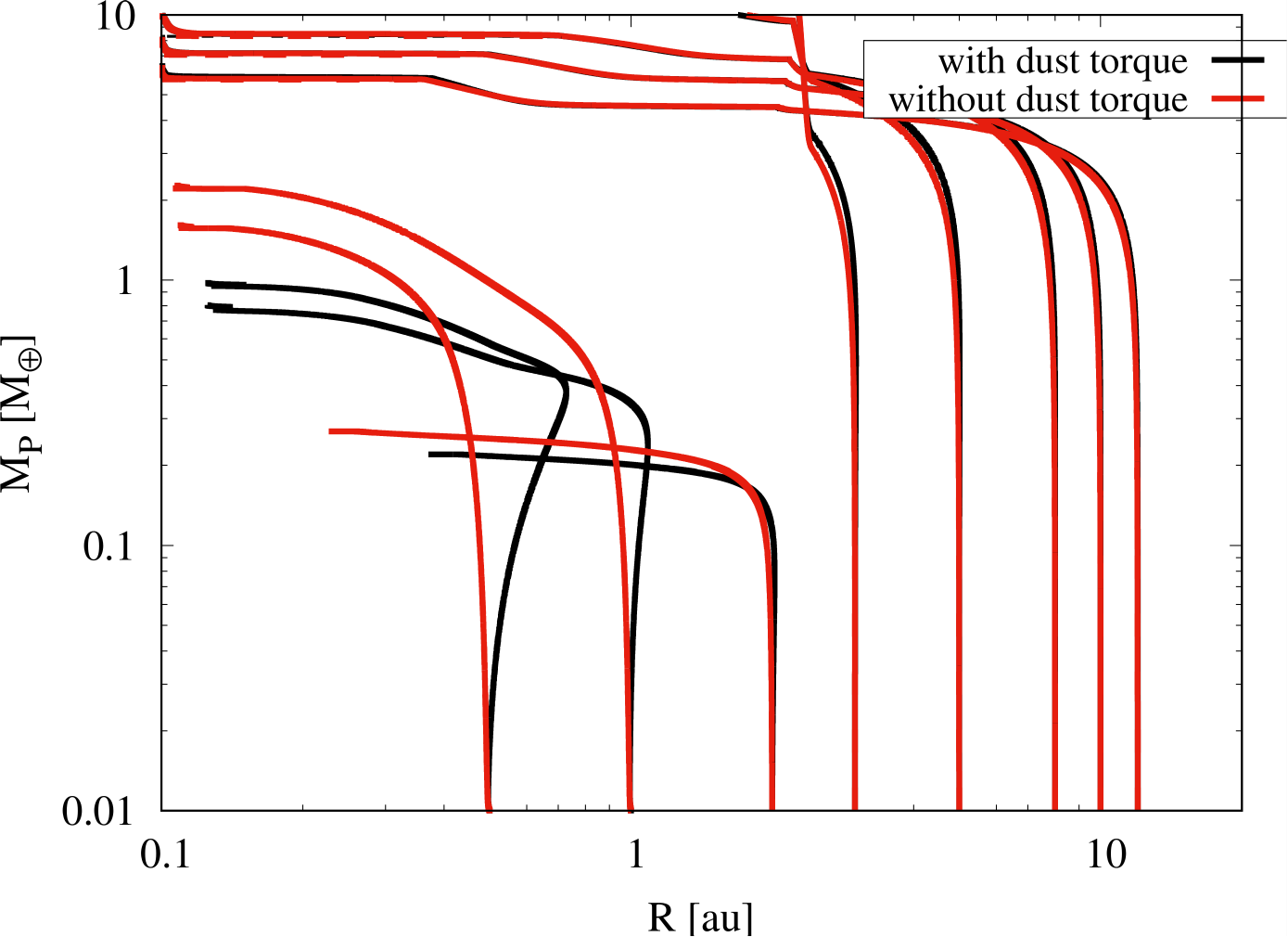} 
    \caption{Planet formation tracks (planet total mass vs. semi-major axis) for the fiducial disk. Planets inside 2~au begin their formations inside the water ice-line. The dashed lines in panel b) represent the masses of planet's core.} 
    \label{fig1_sec3.2}
\end{figure}

In Figure~\ref{fig1_sec3.2}, we illustrate planet formation tracks in the fiducial disk model.  To analyze the impact of the dust component on the migration history, we calculate planet formation tracks that incorporate both gas and dust components (represented by black lines), as well as tracks that consider only the gas component (shown in red). Two distinct behaviors emerge based on whether the planets begin their evolution within or beyond the water ice-line, initially located at 2.25~au.

{\it Migration from Inside the Water Ice-line} \/--
The evolution of planets that begin at distances of 0.5 au and 1 au is greatly influenced by the dust torque. In both scenarios, the planets experience outward migration, resulting in final masses that are lower than those observed in a dust-free disk (see below). This occurs because, although the pebble mass-weighted mean Stokes numbers are lower inside the water ice-line, the rising dust-to-gas ratio generates positive total torques at smaller masses due to the impact of the positive dust torque
\citep[see][]{Guilera+2023}.

\begin{figure}
    \centering
    \includegraphics[angle= 270, width=\columnwidth]{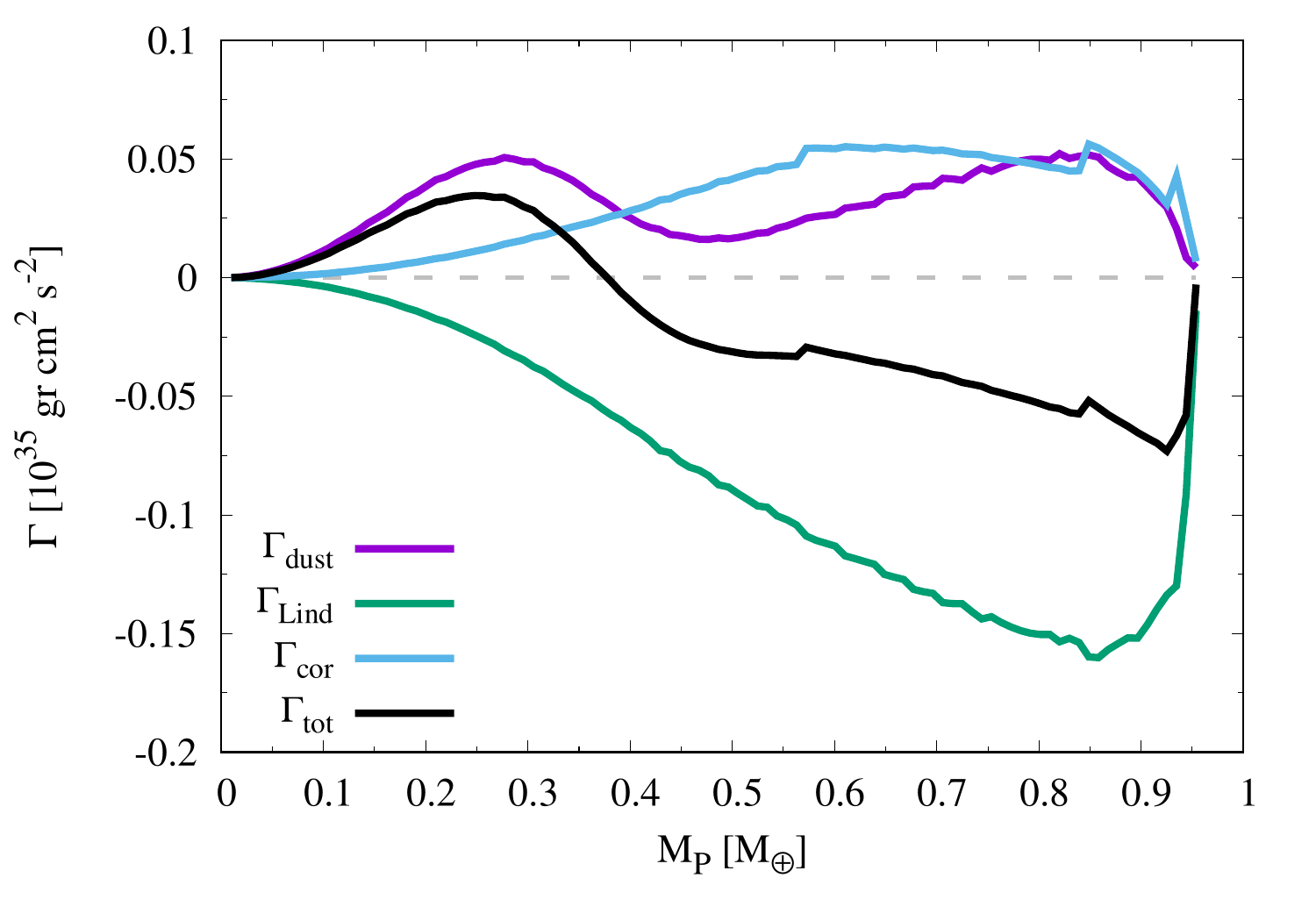}
    \caption{Individual torque components and the total torque on the planet, as a function of the planet mass for the planet initially located at 0.5~au in the fiducial run.}
    \label{fig2_sec3.2}
\end{figure}

This is illustrated in Figure~\ref{fig2_sec3.2}, where we show the torque components and the overall torque experienced by the planet as a function of its mass, for the planet initially positioned at 0.5 au. When factoring in the dust torque, the total torque stays positive until the planet attains a mass of approximately $\sim 0.4~M_{\oplus}$. Beyond this mass, the Lindblad torque, which scales with the planet's mass, starts to dominate resulting in a negative total torque. Note all the components of the total torque vanish  at the end of the simulation because the disk is dispersed.   

{\it Migration from Outside the Water Ice-line} \/--
For planets initially located outside the water ice-line the effects of the dust torque are not significant. This is due to two different reasons. On the one hand, the dust-to-gas mass ratio decreases rapidly beyond the water ice-line, leading to a rapid decline of the positive dust torque\footnote{The dust torque is always positive in the gas-dominated regime for $\text{St} \lesssim 0.2$ \citep[see][]{Guilera+2023}.} On the other hand, as we show in  \citet{Guilera+2023}, when pebble accretion rates are high, and the formation time-scales are shorter than the migration time-scales, planets grow essentially in-situ until they reach a few Earth masses. This agrees with the results presented in \citet{Guilera2021}, who found that planets growing by pebble accretion beyond the water ice-line have formation times-scales of $\sim 10^5$~yr. The planet initially located at 2~au grows faster than those further out but its evolution is rather insensitive to the presence of dust since the dust-to-gas mass ratio decreases rapidly enough just inside the water ice-line.

\begin{figure}
    \centering    \includegraphics[width=\columnwidth]{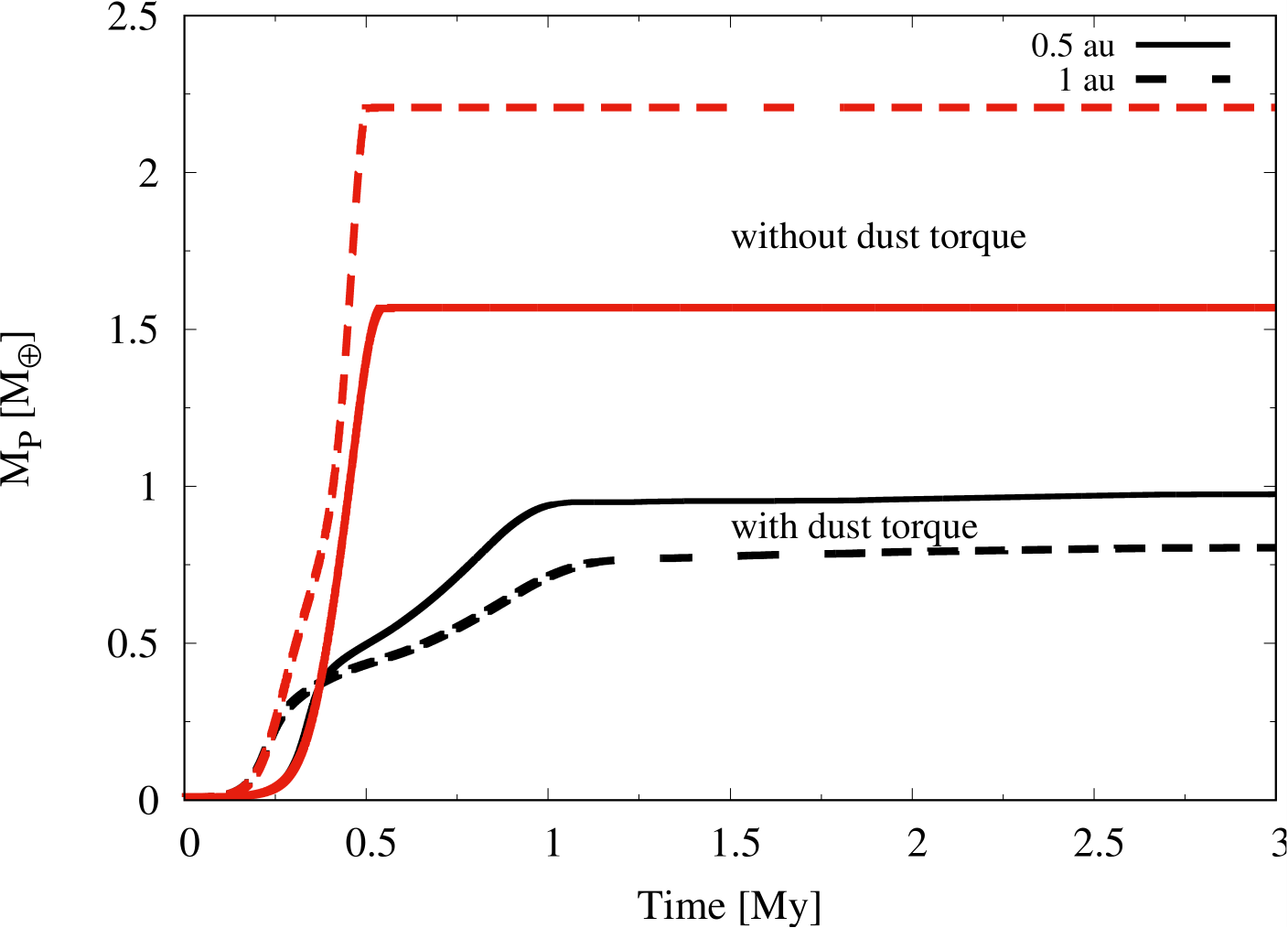}
    \caption{Planet mass versus time for the planets initially located at 0.5 and 1~au (solid and dashed lines). The black/red lines indicate whether the dust torque is considered/ignored. } 
    \label{fig3_sec3.2}
\end{figure}

{\it Final Masses and Formation Time-scales} \/--
Figure~\ref{fig3_sec3.2} illustrates the impact of the dust torque on both the final masses for the planets and their formation time-scales. In scenarios without dust torque (the red lines), the formation time-scales for planets situated at 0.5 au (the solid line) and 1 au (the dashed lines) are approximately 0.5 Myr for both cases. In contrast, when the dust torque is taken into account (the black lines), the formation time-scales increase to about 1.25 Myr for planets at 0.5 au and 1.5 Myr for those at 1 au.

\begin{figure}
    \centering    \includegraphics[width=\columnwidth]{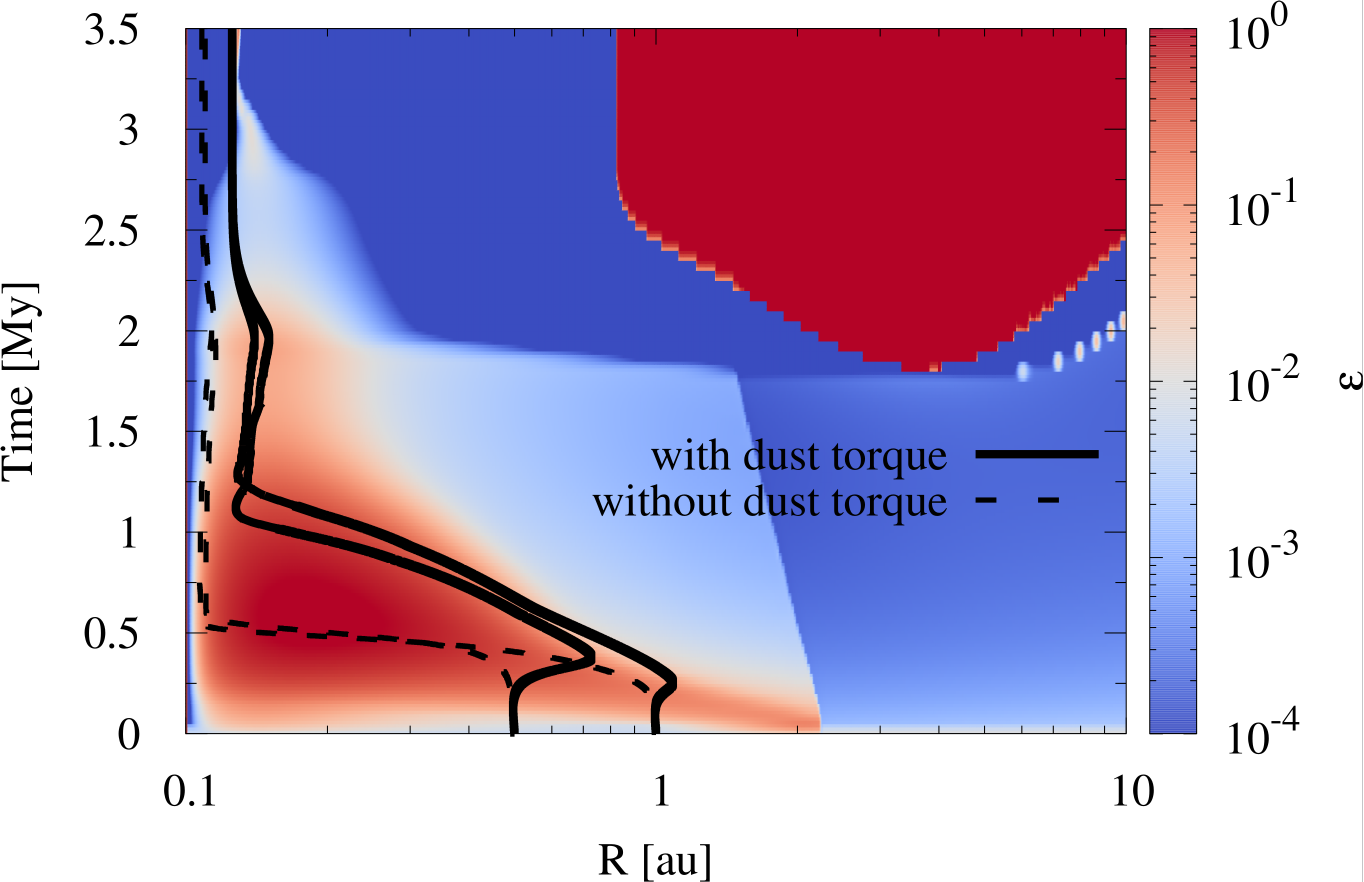}
    \caption{Planet formation tracks for the planets initialed located at 0.5~au and 1~au overlaid on the time evolution map of the dust-to-gas ratio, $\epsilon= \Sigma_{\text{d}}/\Sigma_{\text{g}}$.} 
    \label{fig4_sec3.2}
\end{figure}

Figure~\ref{fig4_sec3.2} illustrates why the planets initially positioned at 0.5 and 1 au end up with lower masses when accounting for the effects of dust torque. In the absence of the dust torque (red lines), planets grow and continuously migrate through a region where the dust-to-gas ratio is highest; significantly boosting the pebble accretion rates. However, when the dust torque is taken into account (black lines), both planets initially migrate outward to a region with a reduced dust-to-gas mass ratio, resulting in less efficient growth.

\subsection{Planet Formation Tracks in Massive Disks}
\label{sec_massive}

\begin{figure}
    \centering
    \includegraphics[width=\columnwidth]{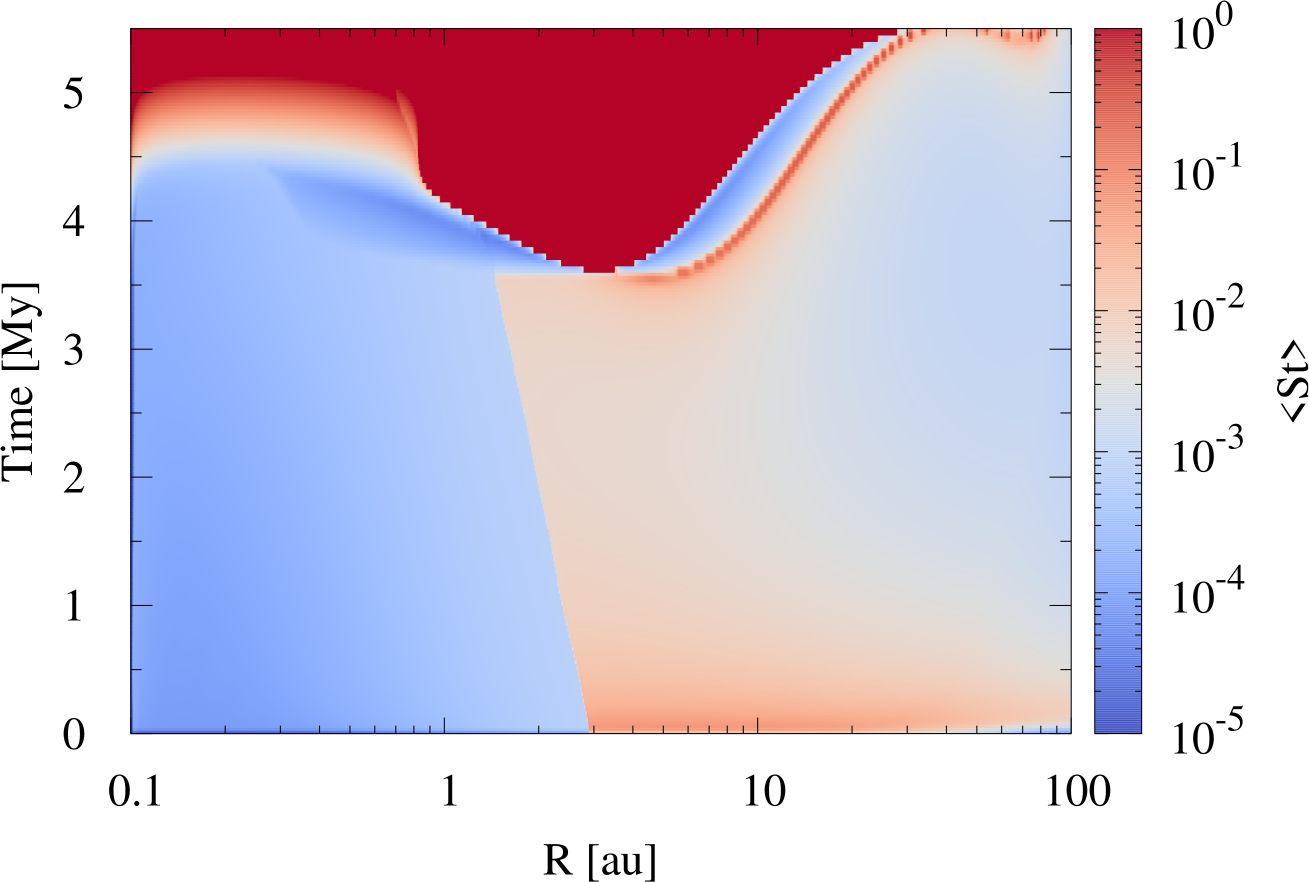} \\
    \centering
    \includegraphics[width=\columnwidth]{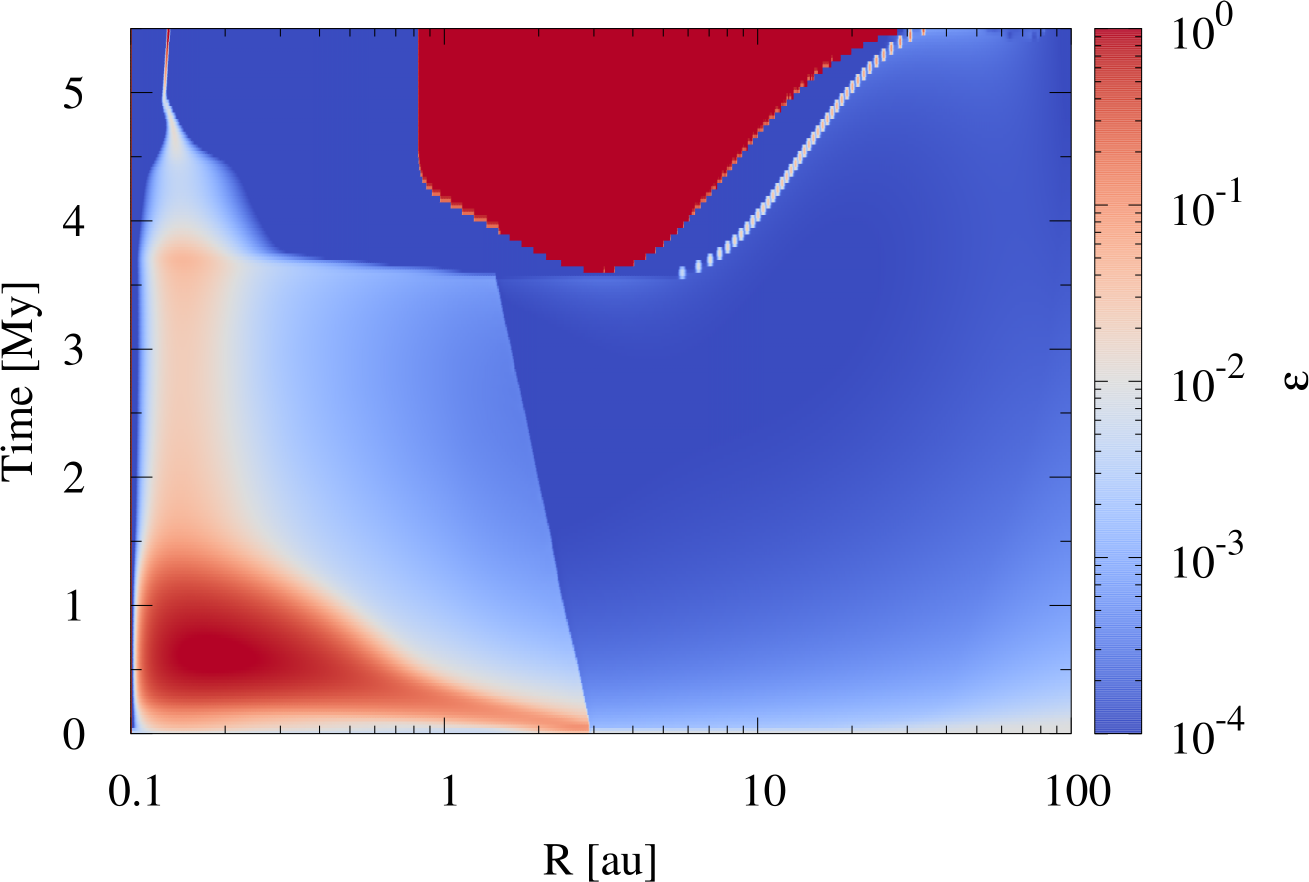}
    \caption{Time evolution maps of the particle mass-weighted mean Stokes numbers (top) and the dust-to-gas mass ratio $\epsilon= \Sigma_{\text{d}}/\Sigma_{\text{g}}$ (bottom) for the case of a massive disk with $M_{\rm disk} = 0.1M_{\odot}$.} 
    \label{fig1_sec3.3.1}
\end{figure}

In Figure~\ref{fig1_sec3.3.1}, we present the time evolution maps of the particle mass-weighted mean Stokes numbers of the dust distribution (top panel) and the dust-to-gas mass ratio (bottom panel) for a massive disk with an initial mass of $0.1~M_{\odot}$ solar masses. The time evolution in each of these panels closely resembles their counterparts in Figure~\ref{fig1_sec3.1} because of a compensatory effect. Although dust can grow to slightly larger sizes — limited by fragmentation in the inner disk, which depends on gas surface density, and by drift in the outer disk, which depends on dust surface density \citep[e.g.,][]{Birnstiel12, Drazkowska16, Guilera2021, Birnstiel2024} — the Stokes numbers are directly proportional to dust sizes but inversely proportional to gas surface density, which is higher in this scenario. As a result, there are no significant differences in mass-weighted mean Stokes numbers across the disk between the moderate-mass and massive disk cases. Notably, the main differences compared to the fiducial case include the opening of the gas gap at approximately 3.5 Myr and the initial location of the water ice-line at around 3 au.

\begin{figure}
    \centering
    \includegraphics[width=\columnwidth]{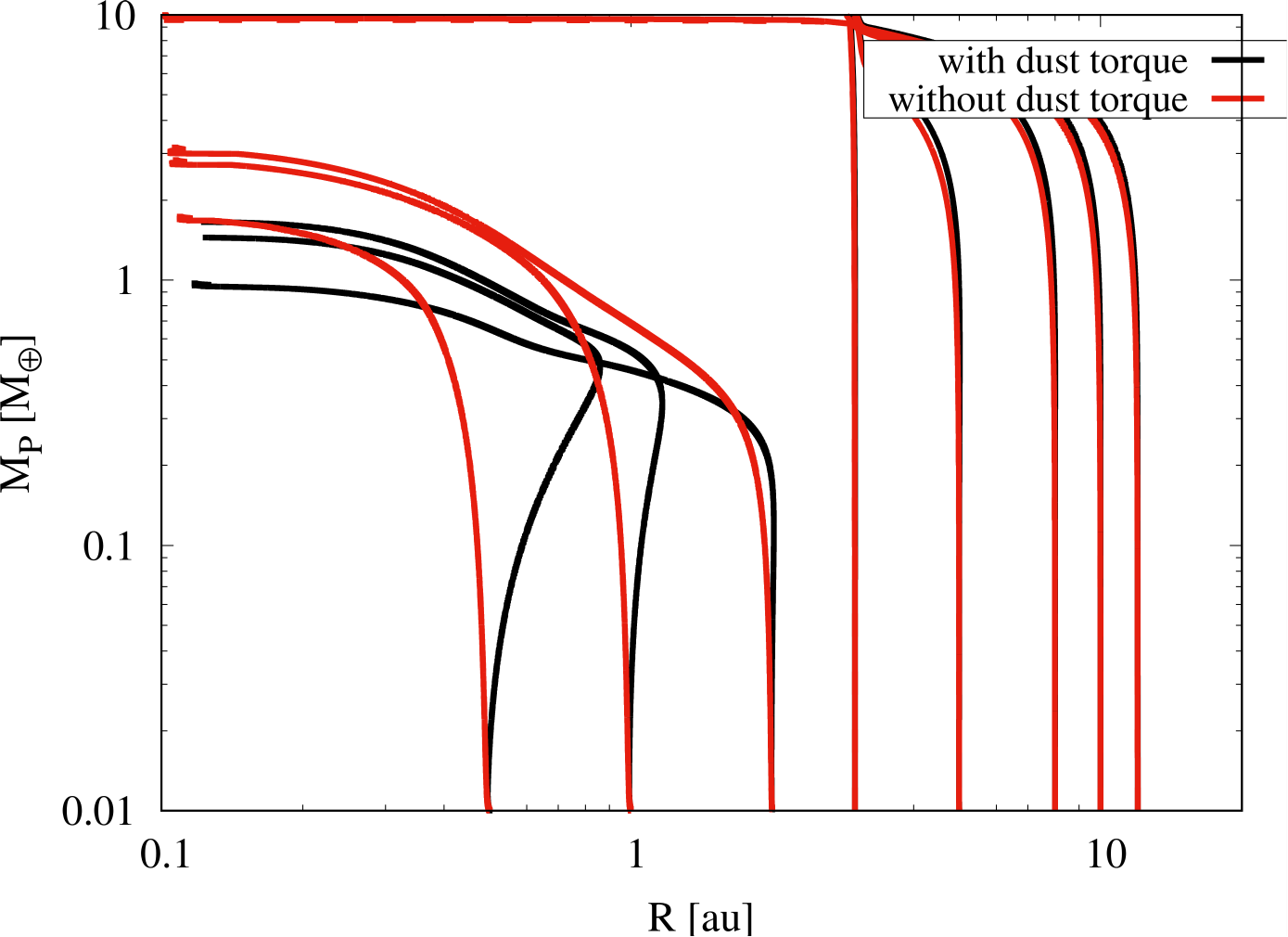} 
    \caption{Same as Figure~\ref{fig1_sec3.2} but adopting an initial disk mass of $M_{\rm disk} = 0.1M_{\odot}$.} 
    \label{fig2_sec3.3.1}
\end{figure}

In Figure~\ref{fig2_sec3.3.1}, we display the planet formation tracks for the case of the massive disk. Once again, for planets located beyond the water ice-line (i.e., those initially positioned beyond 3 au), the dust torque is not significant in spite of the higher Stokes numbers of pebbles in that region. This is primarily due to the rapid formation timescales of these planets and the swift decline of the dust-to-gas ratio in that region of the disk \citep[see][]{Guilera+2023}. Similar to the fiducial disk model, for planets situated inside the water ice-line the dust torque plays a crucial role. In all scenarios, the dust torque dominates the total torque during the early stages, causing planets (especially those starting at 0.5 au and 1 au) to migrate outwards. The dust torque significantly alters the planet formation tracks in all cases, resulting in planets that are less massive compared to scenarios where the dust torque is not considered. As demonstrated by \citet{Guilera+2023}, the dust torque is more effective due to the increased dust-to-gas ratio (see Figure~\ref{fig1_sec3.3.1}), even with the relatively low Stokes numbers inside the water ice-line. Additionally, as noted in the previous section, the final planet masses decrease because the dust torque pushes planets to migrate outward into areas with lower dust-to-gas ratios, which in turn reduces the rates of pebble accretion.

\subsection{Planet Formation Tracks in Metal-rich Disks}
\label{sec_metal-rich}

\begin{figure}
    \centering
    \includegraphics[width=\columnwidth]{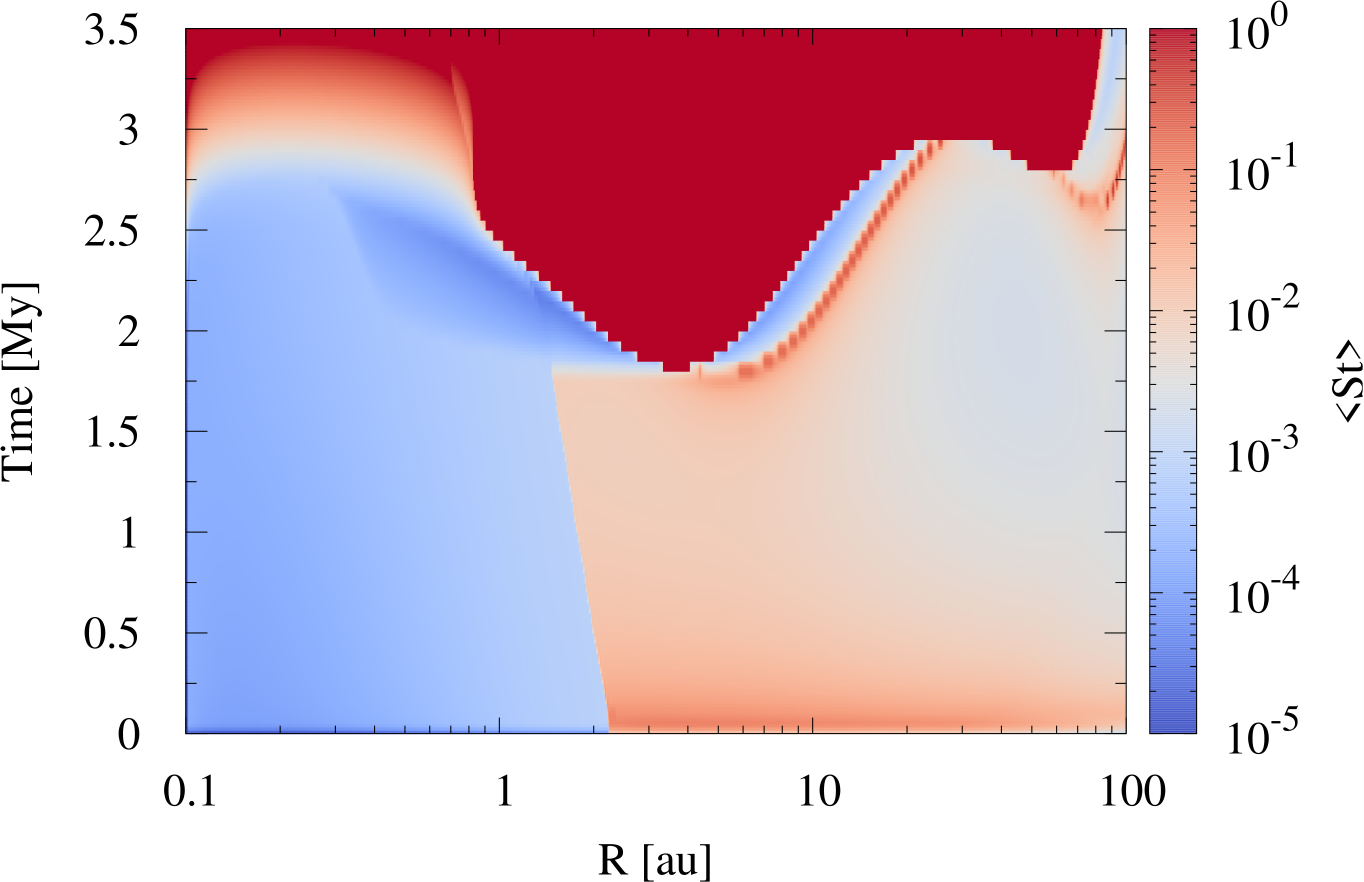} \\
    \centering
    \includegraphics[width=\columnwidth]{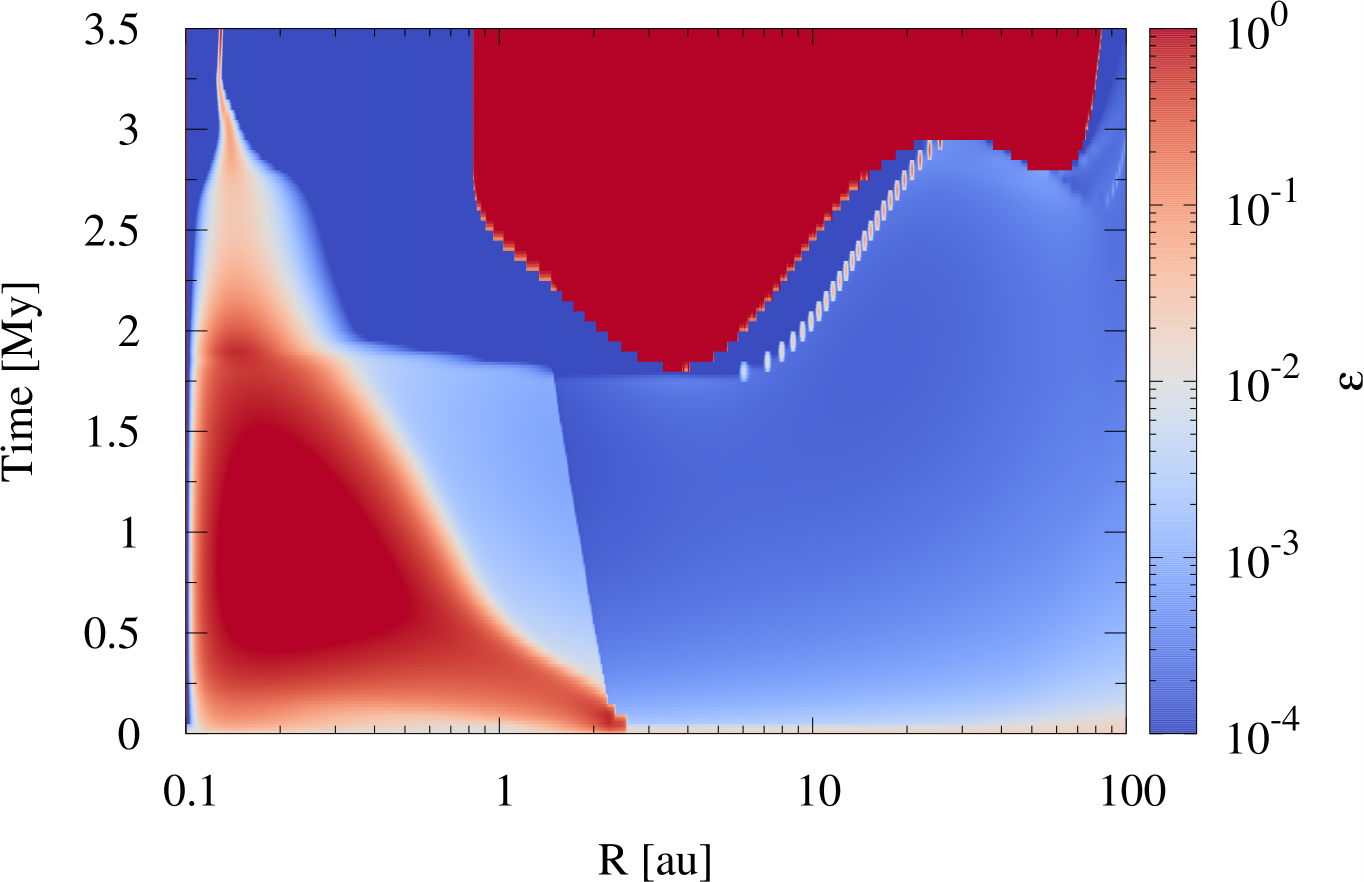}
\caption{Time evolution maps of the particle mass-weighted mean Stokes numbers (top) and the dust-to-gas mass ratio $\epsilon= \Sigma_{\text{d}}/\Sigma_{\text{g}}$ (bottom) for the case of a metallic disk with an initial dust-to-gas ratio $\epsilon=0.02$.}     
    \label{fig1_sec3.3.2}
\end{figure}

We investigate next the impact of the initial disk metallicity. Figure~\ref{fig1_sec3.3.2} presents the time evolution maps of the particle mass-weighted mean Stokes numbers of the dust distribution (top panel) and the dust-to-gas mass ratio (bottom panel). As observed in the previous section, the evolution of these profiles is akin to that of the fiducial case. However, with a higher initial disk metallicity, the dust-to-gas ratio in this scenario is significantly elevated compared to the previous cases.

\begin{figure}[h!]
    \centering
    \includegraphics[width=\columnwidth]{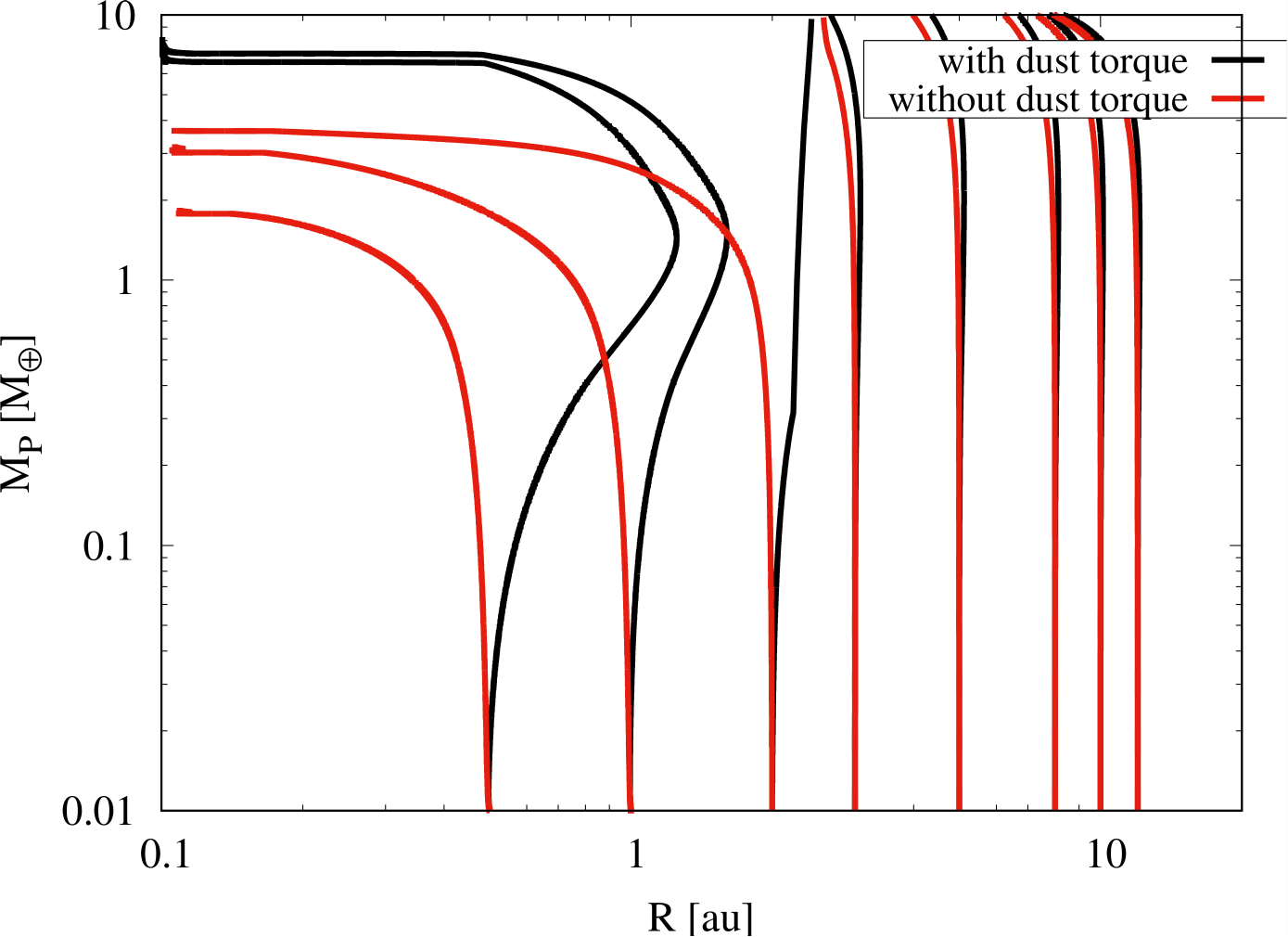} \\
    \centering
    \includegraphics[width=\columnwidth]{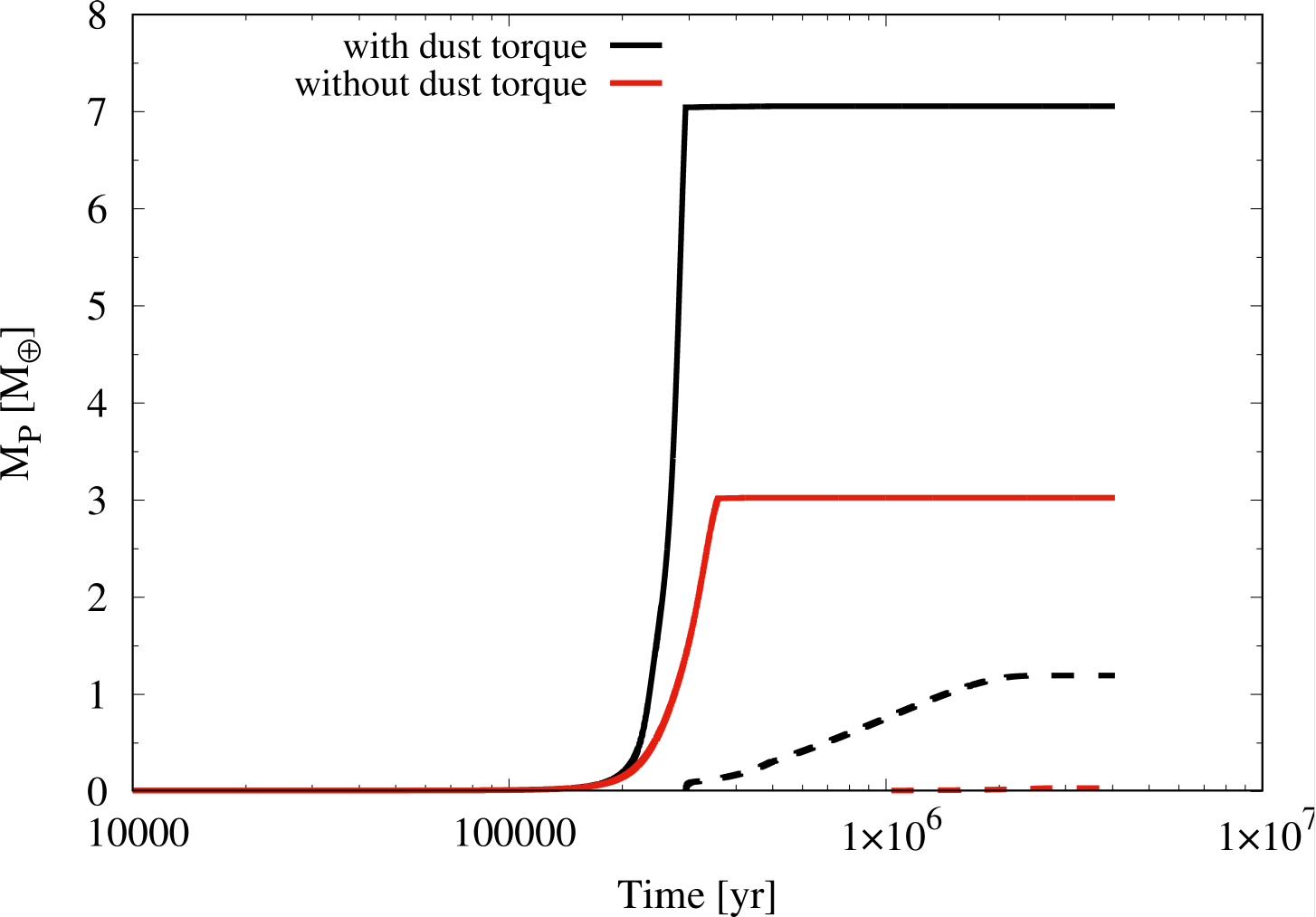}
    \caption{Top panel: same as Figure~\ref{fig1_sec3.2} but adopting an initial dust-to-gas ratio $\epsilon= 0.02$. Bottom panel: time evolution of the core mass (solid lines) and envelope mass (dashed lines) for the planet initially located at 1~au.} 
    \label{fig2_sec3.3.2}
\end{figure}

\begin{figure}
    \centering    \includegraphics[width=\columnwidth]{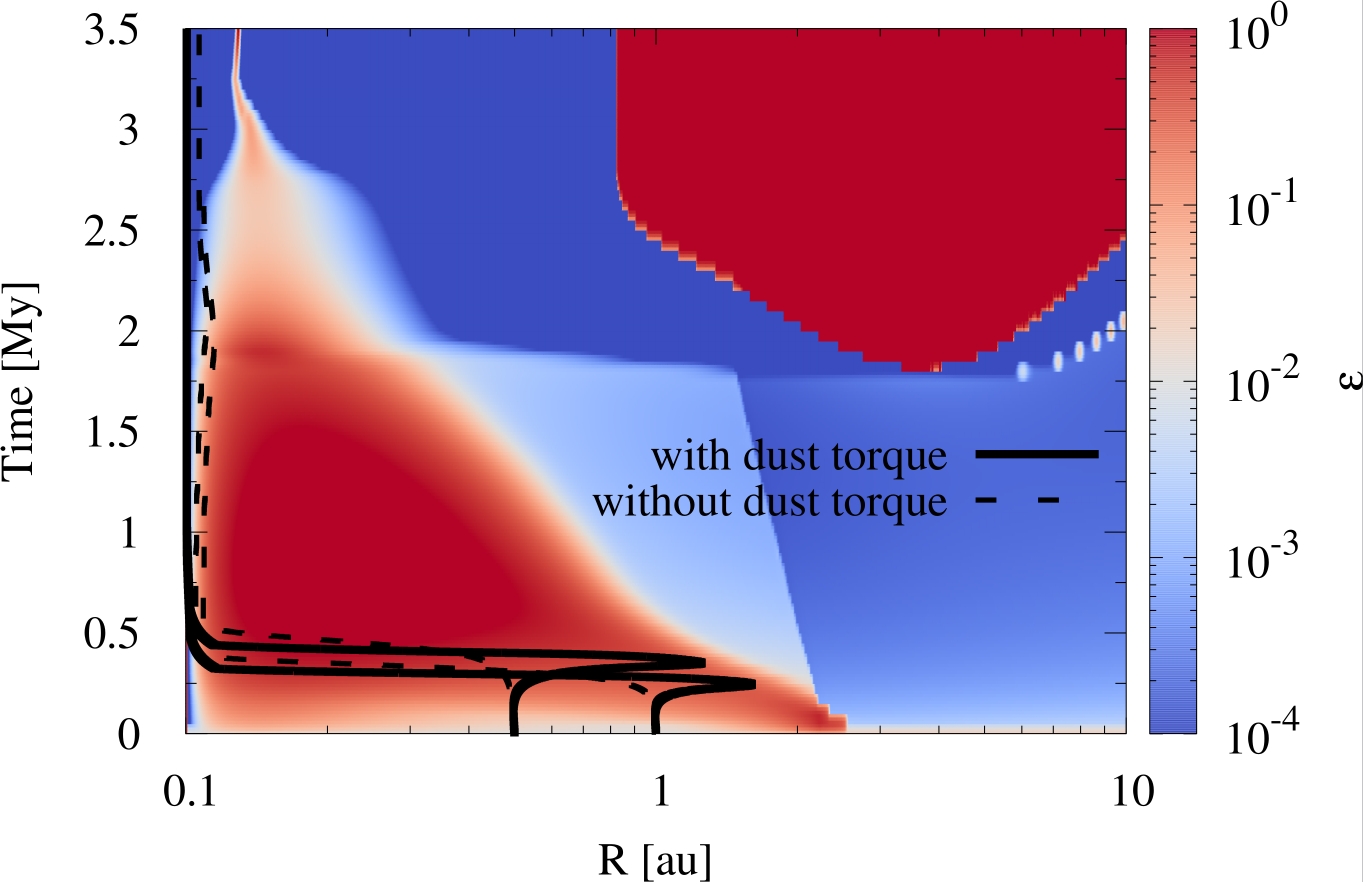} \\
    \centering    \includegraphics[width=\columnwidth]{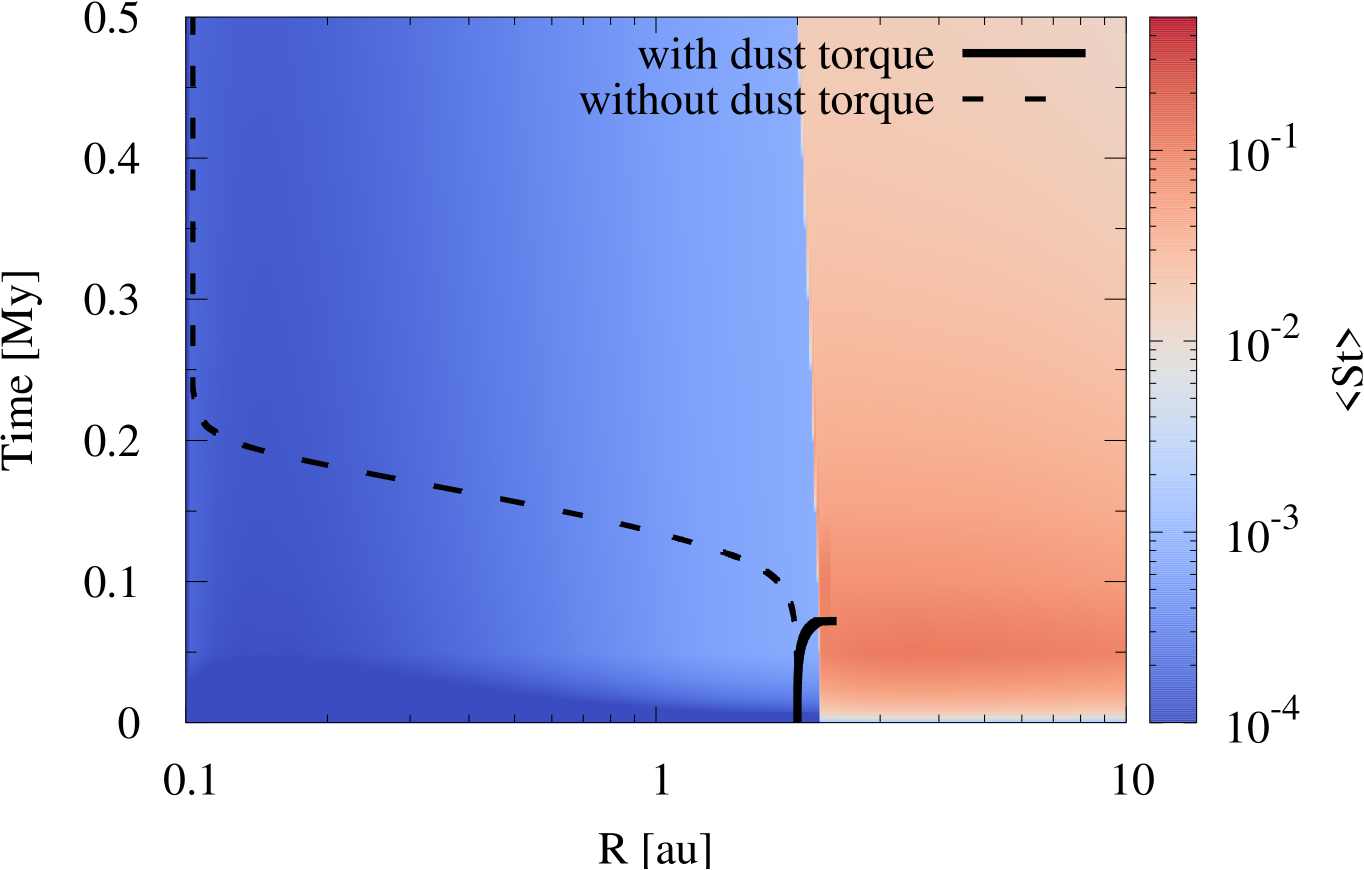} 
    \caption{Top panel: planet formation tracks for the planets initialed located at 0.5~au and 1~au overlaid on the time evolution map of the dust-to-gas ratio ($\epsilon= \Sigma_{\text{d}}/\Sigma_{\text{g}}$). Bottom panel:  planet formation track for the planet initialed located at 2~au overlaid on the time evolution map for the particle mass-weighted mean Stokes numbers. To improve the visualization, only the evolution until 0.5~Myr is shown.} 
    \label{fig3_sec3.3.2}
\end{figure}

This increased dust-to-gas ratio substantially affects the formation tracks of planets initiating their formation inside the water ice-line (specifically, those at 0.5, 1, and 2 au, as shown in the top panel of Figure~\ref{fig2_sec3.3.2}). In these instances, when the dust torque is accounted for (black lines), all these planets migrate outward during the early stages of their development. Unlike the previous cases, these outward-migrating planets ultimately attain larger masses than they would have reached had the dust torque been ignored. Notably, the planet starting at 2 au crosses the water ice-line, sparking rapid formation through the accretion of pebbles with larger Stokes numbers (as highlighted in the bottom panel of Figure~\ref{fig3_sec3.3.2}). 

The planets initially located at 0.5 au and 1 au achieve relatively massive cores, enabling them to attract a significant envelope between $\sim 0.5$ -- $1~\text{M}_{\oplus}$. Unlike the fiducial case, these planets remain in regions of high dust-to-gas ratios as they migrate outward (see the top panel of Figure~\ref{fig3_sec3.3.2}). In contrast, when the dust torque is excluded, planets that initiate their formation inside the water ice-line end with negligible envelopes due to their lower mass cores (as illustrated in the bottom panel of Figure~\ref{fig2_sec3.3.2}).

Finally, for planets that are initially situated beyond the water ice-line (those positioned beyond 3 au in Figure~\ref{fig2_sec3.3.2}), we once again observe that the dust torque has minimal impact. The dust torque only has a minor effect on delaying inward migration, leading to planet formation paths that remain mostly the same.

\section{The Impact of Pebble Accretion}
\label{sec_4}

The first attempt to incorporate pebble accretion in the context of pebble-driven torques was made by \citet{Regaly2020}. In their multi-fluid model, pebble accretion was implemented by artificially reducing the pebble surface density, while maintaining a fixed accretion radius and treating the accretion efficiency as a variable parameter. This approach effectively sidestepped self-consistent modeling of pebble trajectories converging at the planet's location. In spite of this simplification, \citet{Regaly2020} demonstrated that for low Stokes numbers and low-mass planets, including pebble accretion can lead to a significantly strong positive dust torque. A similar finding was reported more recently by \citet{Chrenko+2024}, who investigated the effect of incorporating pebble accretion on the dust torque using a new hybrid method that allows pebble trajectories to converge to the planet's radius (as shown in their Figures 2 and 3). Furthermore, through their Equations~(16) to (18), \citet{Chrenko+2024} provided a relatively straightforward recipe to fit their results. 

\begin{figure}
    \centering
    \includegraphics[width=\columnwidth]{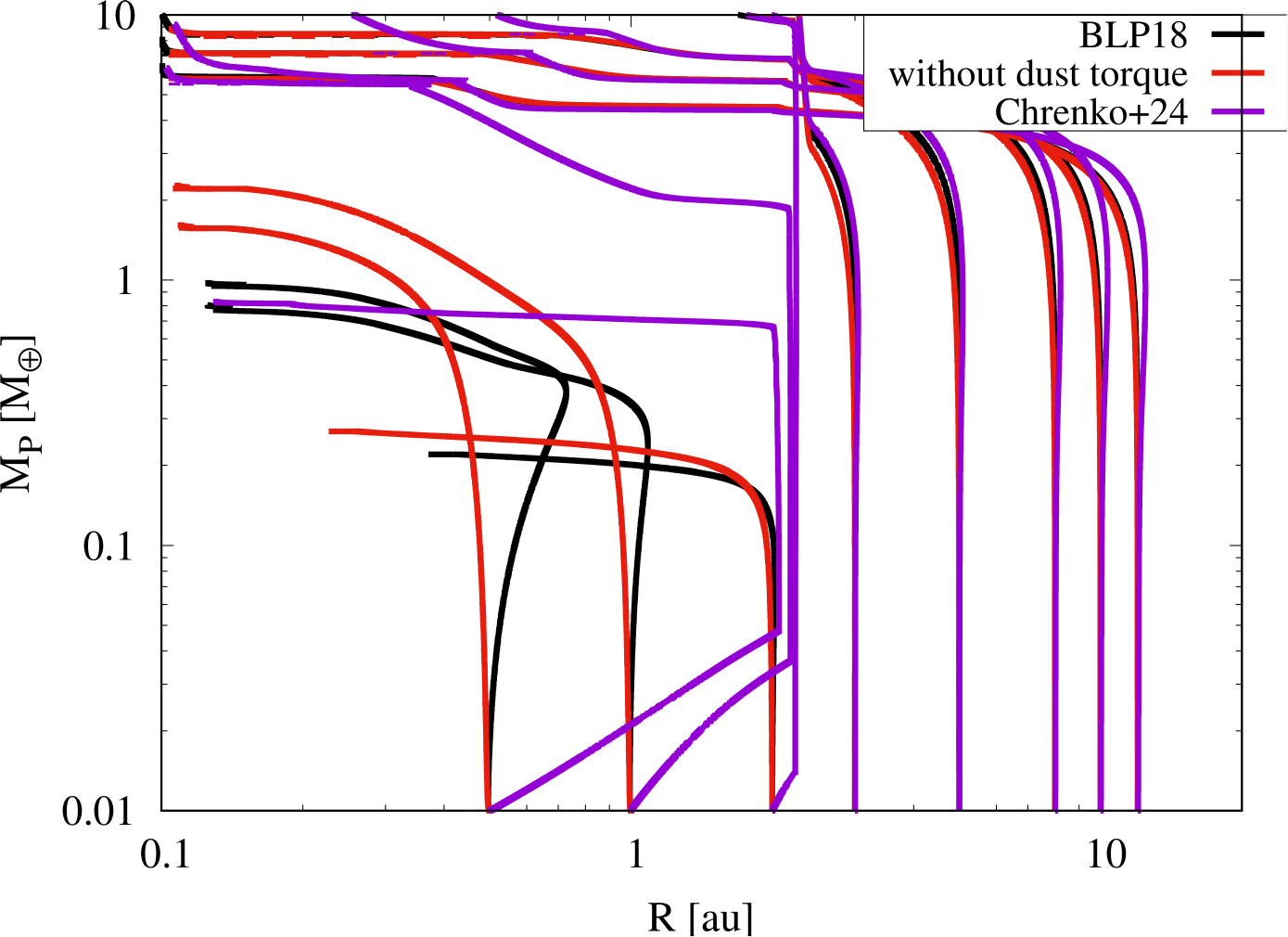}
    \caption{Planet formation tracks for the fiducial disk --as in Figure~\ref{fig1_sec3.2}-- including in this case the impact of pebble accretion onto the dust torque (violet lines).} 
    \label{fig1_sec4}
\end{figure}

It is interesting to compare the planet formation tracks generated using the results from BLP18 and those from \citet{Chrenko+2024}, who used an analytical formula to fit their numerical results. In Figure~\ref{fig1_sec4}, we present the planet formation tracks for the fiducial case, contrasting scenarios that do not consider the dust torque (red lines) with those that incorporate results from BLP18 (black solid lines) and the tracks obtained by implementing in our code the new recipes provided in \citet{Chrenko+2024} (violet solid lines). For planets initially located beyond the water ice-line (i.e., beyond 3 au), Figure~\ref{fig1_sec4} illustrates that, although the overall formation tracks remain similar, the dust torque appears more effective at larger distances when employing the prescriptions of \citet{Chrenko+2024}. This results in a slightly more pronounced delay of rapid inward migration for those planets. In contrast, for planets that begin their formation inside the water ice-line, the effects of the dust torque become markedly more significant. This is because, according to the results in \citet{Chrenko+2024} the positive magnitude of the dust torque is notably larger for low-mass planets and small Stokes numbers when pebble accretion is taken into account. Consequently, these planets migrate outward significantly during the early stages of formation. In addition, in this case all the inner planets cross the water-ice line and their formation is influenced by the accretion of ice-rich pebbles with larger Stokes numbers (as in the case of the metallic disk). Eventually, at some point along their formation pathways, the gas torque begins to dominate, leading the planets to migrate inward. The fact that the dust torque is more efficient for inner planets aligns well with the estimates provided by \citet{Chrenko+2024}, who employed a uniform pebble flux (see their Figure 7). 

\section{Summary and Discussion}
\label{sec_5}

The studies by BLP18, \citet{Regaly2020}, and, more recently by \citet{Chrenko+2024}, have demonstrated that the asymmetric distribution of pebbles surrounding a planet can exert a significant torque on it, even though solid materials make up only a small fraction of the protoplanetary disk's mass. Building on these findings \citep[as well as on our previous work,][]{Guilera+2023}, in this study we have quantified for the first time the influence of dust torque on the migration of planets growing through pebble accretion until either the disk dissipates or, conservatively, the planets reach a mass of $10~M_{\oplus}$. We accomplished this goal with PLANETALP, a comprehensive framework for modeling the formation history of planets that incorporates dust growth and evolution \citep{Ronco2017, Guilera2017b, Guilera2019, Venturini20ST, Guilera2021}. 
    
We first employed a fiducial disk model with a moderate mass of $0.05M_{\odot}$ and a low $\alpha$-viscosity parameter of $10^{-4}$. Previous works on dust growth and evolution suggest that the efficiency of planet formation via pebble accretion is highly sensitive to the level of turbulence in the disk \citep[e.g.,][]{Venturini20ST, Venturini20Letter, Drazkowska21, Guilera2021}. On observational grounds, ALMA data suggest low $\alpha$-viscosity values \citep[e.g.,][]{Dullemond2018, Flaherty2020}.  Therefore, based on these results and the findings of \citet{Venturini20Letter} and \citet{Guilera2021}, we considered a low $\alpha$-viscosity value, which enhances planetary formation efficiency. 

{\it Impact of Dust Torque in Fiducial Disk Models} \/--
Our findings for the fiducial disk model can be summarized as follows. Due to the accumulation of dust in the inner regions of the disk, driven by dust drift, the dust-to-gas mass ratio increases significantly there, making the dust torque a dominant component of the total torque, even for low Stokes numbers \citep[as shown in][]{Guilera+2023}. This leads to outward migration for the innermost planets at early times. These planets move outward to a region where the dust-to-gas ratio diminishes, resulting in a decreased rate of pebble accretion, which causes them to end up with lower masses compared to scenarios where dust torque is not considered. In contrast, the dust torque does not significantly affect planets initially located beyond the water ice-line, primarily due to their relatively rapid formation times and the quick decline in the dust-to-gas ratio resulting from dust radial drift. As shown in \citet{Guilera+2023}, this decay considerably reduces the magnitude of the dust torque.

{\it Impact of Dust Torque in Massive and Metal Rich Disks} \/--
We also investigated the role of the dust torque in the migration of planets within disks characterized by higher masses and metallicities. In both scenarios, we found that the mass-weighted mean pebble Stokes numbers across the disk remained comparable to those in the fiducial case. This similarity arises from the fact that dust growth is limited by fragmentation in the inner disk and by radial drift in the outer regions, resulting in minimal differences between these models. Consequently, the outcomes for these scenarios are qualitatively similar to the fiducial case. Although the dust torque plays a crucial role in the formation of planets initially located inside the water ice-line, it is less significant for those that start to form beyond the ice-line. This is due to the increased dust-to-gas mass ratio inside the water ice-line and the corresponding decrease of it outside, as a consequence of dust radial drift. 

Notably, the disk with an initial metallicity of $Z = 0.02$ exhibits a higher dust-to-gas ratio, resulting in a more pronounced impact of the dust torque. In this case, planets starting at 0.5 au and 1 au migrate outward until they reach approximately $2~M_{\oplus}$, while the planet initially at 2 au crosses the water ice-line. This change not only affects the final mass of the planet but also alters its composition. A similar trend was observed in \citet{Guilera2021} when updated prescriptions for the thermal torque were implemented. 

Given that recent estimates of the initial solar metallicity range between approximately $Z_\odot \simeq$ 0.015 and 0.019 \citep[e.g.,][]{vonS2016, BorexinoNeutrinos2020, Lodder2020, Asplund2021}, which may well be representative of the initial dust-to-gas mass ratio in the proto-solar nebula, we argue that the dust torque could be a significant factor in models that account for the formation of terrestrial planets in our solar system via pebble accretion \citep[e.g.,][]{Morby15, Johansen2021}, as well as in general models of terrestrial-like planet formation based on pebble accretion \citep[e.g.,][]{Lambrechts19, Ogihara2020, Venturini20ST, Izidoro2021}.

{\it The Dynamical Impact of Pebble Accretion via the Dust Torque} \/--
The model for the dust torque built by 
\citet{Guilera+2023}, based on the results of BLP18, does not account for pebble accretion. We examined the impact of this process on the migration history of growing planets by adopting the prescriptions recently proposed by  \citet{Chrenko+2024}. 

For planets initiating their formation within the water ice line, the influence of dust torque becomes more pronounced compared to the findings from BLP18. This is primarily because, as \citet{Chrenko+2024} demonstrated, the dust torque tends to increase for low-mass planets with low Stokes numbers when pebble accretion is taken into account.

For planets initially positioned beyond the water ice line, we observe that considering the effects of pebble accretion on the dust torque has minimal influence on the formation tracks. This is even though the dust torque incorporating accretion is larger in magnitude than the one calculated using BLP18's results. This occurs primarily because of the brief timescales for planet formation and the rapid decrease in the dust-to-gas mass ratio beyond the water-ice line. Furthermore, incorporating solid accretion enhances the magnitude of the dust torque at larger distances.  Moreover, we note that the impact of the dust torque, when including accretion, seems to increase at larger distances from the star. This enhancement might stem from the different dependence of the dust torque on the normalized torque parameters, contrasting our approach (using the results from BLP18)  with those from \citet[][see their discussion in Sec. 3.5]{Chrenko+2024}.

We note that we adopted a conservative approach by not extrapolating beyond the dust torque map computed by BLP18, or the prescriptions provided by \citet{Chrenko+2024}. While the values for the pebble Stokes number beyond the water ice-line generally fall within the ranges used by BLP18 and \citet{Chrenko+2024}, those inside the water ice-line are consistently lower than the minimum value in the grid. Therefore, extending the previous map to incorporate lower Stokes numbers (and smaller planet masses) and developing comprehensive analytical prescriptions that accurately represent detailed hydrodynamical simulations, such as those conducted by \citet{Chrenko+2024}, would be beneficial. 
Nevertheless, based on our conservative approach, we expect the dust torque to be significant for lower Stokes numbers and planet masses. 

We also note that the planets we modeled start forming at the beginning of the simulations, specifically at $t = 0$~yr. As demonstrated by \citet{Drazkowska21} and \citet{Guilera2021}, if planetary embryos do not start forming in the early phases of disk evolution, pure pebble accretion becomes inefficient due to diminishing pebble flux over time caused by dust radial drift. \citet{Voelkel2022} incorporated a model for planetesimal formation through pebble flux and embryo formation from planetesimal growth in a framework similar to the one we use in this study. Their findings revealed that multiple generations of embryos can form prior to disk dissipation, particularly within the water ice-line, and that earlier generations primarily grow via pebble accretion, while later ones tend to grow through planetesimal accretion. More recently, \citet{Kaufmann+2025} studied the formation of first embryos via planetesimal and pebble accretion in planetesimal rings forming by the streaming instability mechanisims. The key findings indicate that although the formation of planetary embryos from filaments can occur within the disk’s lifetime, it is highly sensitive to the distance from the central star. In the inner regions of the disk, planetesimal rings generated by the streaming instability can rapidly give rise to protoplanetary embryos, whereas in the outer disk, embryo formation requires substantially more time. We plan to incorporate these phenomena in future studies to examine further the impact of the dust torque on different generations of formed planets.

Finally, as we mentioned in our previous work \citep{Guilera+2023}, our results inherently carry the same limitations as those present in the torque computations of BLP18. We note that the numerical simulations employed in BLP18 are 2D, and thus do not account for the vertical structure of the disk. We also note that the back-reaction of dust on the gas (i.e., feedback) is neglected in BLP18 \citep[also in][]{Chrenko+2024}. This assumes that the solid-to-gas density ratio remains low both in the bulk of the disk and in the vicinity of the planet. Nevertheless, \citet{Regaly2025} showed recently that the inclusion of the back-reaction of solid material tends to increase the dust torque, especially for very low-mass planets (of the order of Mars mass). 

Another simplification in BLP18 is the omission of planetary migration, which alters the relative motion between solids and the planet and may significantly affect the asymmetry in the dust distribution near the planet. This, in turn, could either enhance or reduce the resultant torque. If migration induces eccentricity in the planet's orbit, the torque behavior may change further, underscoring the need for additional hydrodynamical studies to explore these effects.

\section{Conclusions}
\label{sec_6}

In this study, we examine the impact of the dust torque on the migration of planets forming via pebble accretion. Our model includes a comprehensive framework accounting for disk evolution, dust growth and evolution, and the processes of planet formation and migration, incorporating dust torque values derived from the maps established by BLP18, as well as the recent findings from \citet{Chrenko+2024} incorporating pebble accretion. Based on our model's assumptions, we present the following key findings depending on the initial location of the planetary embryo: \\ 

{\it Planets initially inside the water ice-line} \/-- In this case, the dust torque represents a significant contribution to the total torque due to the increased dust-to-gas mass ratio resulting from dust drift from the outer disk. In fact, at early stages, the dust torque can dominate the total torque acting on the planet, particularly in disks with initial metallicities slightly above the standard 1\% ($\epsilon_0 \gtrsim 0.02$). However, to achieve more accurate results, further detailed simulations extending to lower planet masses and Stokes numbers are necessary, especially when considering the recent results from \citet{Chrenko+2024}.  For these planets, we identify three primary pathways: 

{\it i}) when the outward migration caused by the dust torque drives the planets into regions with a lower dust-to-gas mass ratio—such as in the fiducial and massive disk scenarios—the planets ultimately have lower masses compared to scenarios where only the gas torque is considered; 

{\it ii}) if the dust torque propels the planets outward to areas with a higher dust-to-gas mass ratio, as seen in the metallic disk case, the planets grow more effectively, resulting in larger masses and significant envelopes; 

{\it iii}) when applying the prescriptions from \citet{Chrenko+2024}, all cases show that the planets achieve larger masses. In this scenario, the dust torque plays a more significant role in pushing (torquing) the planets outward, enabling them to cross the water-ice line and facilitating their formation through the accretion of larger pebbles, which also alters the planets' composition by incorporating ice-rich pebbles.

{\it Planets initially beyond the water ice-line} \/--
For these planets, the dust torque does not seem to play a significant role. Although dust may slightly delay inward migration, the planetary formation tracks remain largely unaffected. This can be attributed to the rapid formation timescales of these planets and the swift decline in dust-to-gas mass ratio beyond the water ice-line due to dust drift. 

In conclusion, based on our results, we assert that the dust torque should not be ignored in models of planet formation, especially in research centered on the formation of terrestrial-like planets through pebble accretion. In this regard, the dust torque can significantly influence the overall torque experienced by planets undergoing pebble accretion, thereby impacting their migration patterns and, ultimately, their formation timescale, mass, and composition.

\section{Acknowledgments}
We thank the referee for a useful report with suggestions that helped us improve this manuscript. OMG and M3B are partially supported by PIP 2971 from CONICET (Argentina) and by PICT 2020-03316 from ANPCyT (Argentina). OMG and M3B also thank Juan Ignacio Rodriguez from IALP for the computation managing resources of the Grupo de Astrof\'{\i}sica Planetaria de La Plata. PBL acknowledges support from FONDECYT project 1231205.
The research leading to this work was supported by the Independent Research Fund Denmark via grant ID 10.46540/3103-00205B attributed to MP.  

\appendix

\section{Dust torque from the dust distribution}
\label{appex1}
In this appendix, we compute the dust torque from the dust distribution instead of using only the mass-weighted mean Stokes number of such distribution. Even though it is a more expensive computational approach, it is useful to analyze if there is a significant difference with respect to the simpler approach of considering a mass-weighted mean Stokes number to characterize the dust distribution. To achieve this, we calculate the dust torque at the planet's location for 200 dust species that are evenly spaced logarithmically between $\text{r}_{\text{d}}^{\text{min}}= 1~\mu$m ($\text{St}_{\text{min}}$) and $\text{r}_{\text{d}}^{\text{max}}$ ($\text{St}_{\text{max}}$) \citep[see][for details]{Guilera20}. We note again that in our model -- which is based on the works of \citet{Birnstiel12} and \citet{Drazkowska16} -- most of the mass remains in larger particles. We compute the dust-to-gas ratio corresponding to the i-species given by
\begin{equation}
\epsilon_{\text{i}}= \dfrac{\Sigma_{\text{d}}^{\text{i}}}{\Sigma_{\text{d}}},
\label{eq1-appex1}
\end{equation}
where $\Sigma_{\text{d}}^{\text{i}}$ is computed from the mass dust distribution, and $\sum_{\text{i}} \epsilon_{\text{i}}= \epsilon$, being $\epsilon$ the global dust-to-gas ratio at the location of the planet. Thus, for each $\text{St}_{\text{i}}$ we compute the corresponding dust torque from the torque map calculated in BLP18, $\Gamma_{\text{dust, map}}^{\text{i}}$, and the dust torque for the i-species is  given by
 \begin{equation}
 \Gamma_{\text{d}}^{\text{i}}= \dfrac{\epsilon_{\text{i}}}{\epsilon_0}\Gamma_{\text{d, map}}^{\text{i}}, 
 \label{eq2-appex1}
 \end{equation}  
 with $\epsilon_0= 0.01$. Finally, the total dust torque is given by 
 \begin{equation}
 \Gamma_{\text{d}}= \sum_{\text{i}}\Gamma_{\text{d}}^{\text{i}}.    
 \label{eq3-apeex1}
 \end{equation} 
 
\begin{figure}
    \centering
    \includegraphics[width=\columnwidth]{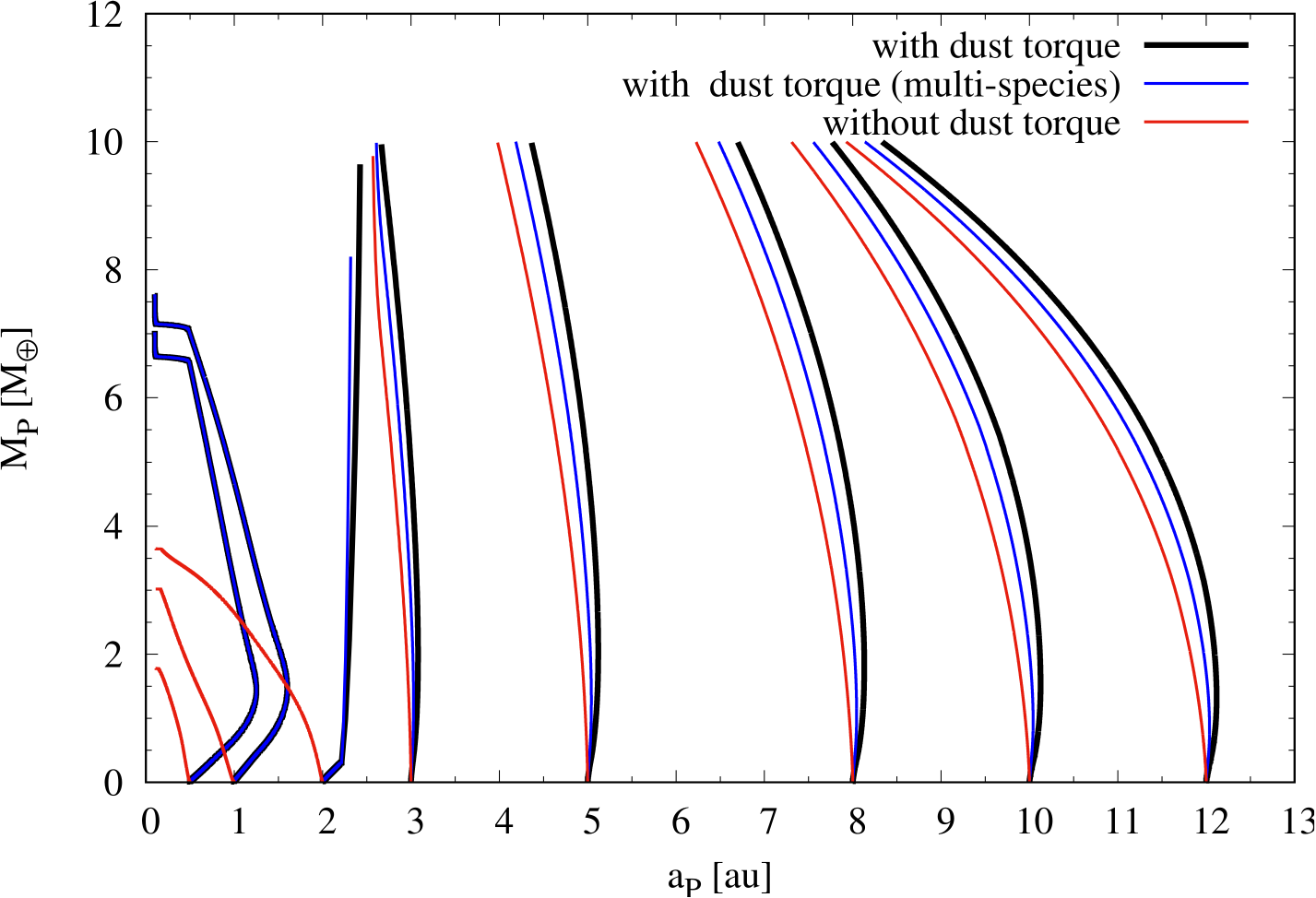}
    \caption{Comparison of the planet formation tracks for the case of an initial disk metallicity of 0.02. The red curves correspond to the case where the total torque is given solely by the gaseous torque \citep[where we use the recipes from][]{jm2017}. The black and blue lines correspond to the case where the dust torque is considered. The black lines illustrate the scenario in which the dust torque is calculated using the mass-weighted mean Stokes number of the dust distribution, whereas the blue lines depict the scenario where the total dust torque is determined by summing the individual torque contributions from the 200 species within the distribution.} 
    \label{fig1_appex1}
\end{figure}

In Figure~\ref{fig1_appex1}, we show a comparison of the planet formation tracks for the case where the dust torque is computed using the mass-weighted mean Stokes number of the dust distribution (the black lines) and the case where the dust torque is calculated for each dust species and then added up to compute the total dust torque (the multi-species case represented by the blue lines). We also plot the case where the dust torque is not considered in the total torque acting on the planets (the red lines). For the planets initially located inside the water ice-line, the planet formation tracks for the cases when the dust torque is considered are identical. This is due to the fact that inside the water ice-line the maximum Stokes number $\text{St}^{\text{max}}$ of the dust distribution is always lower than 0.01 (see the top panel of Figure~\ref{fig1_sec3.3.2}). This also implies that the mass-weighted mean Stokes number is also lower than 0.01. Since we are not extrapolating outside the values of the dust torque map computed in BLP18, for St lower than 0.01 we are using this value to compute the dust torque. Hence, in the multi-species approach the dust torque for each dust species is the same, and the total dust torque is identical to the dust torque computed using the mass-weighted mean Stokes number. On the other hand, beyond the water ice-line the maximum Stokes number $\text{St}^{\text{max}}$ of the dust distribution reaches values larger than 0.01. Thus, in the multi-species approach not all the species contribute with the same value of the dust torque (particles with larger St contribute with larger dust torques). For those planets initially located beyond the water ice-line, we can see that there is a slight difference between the planet formation tracks when the dust torque is considered. Using the mass-weighted mean Stokes number of the dust distribution to calculate the dust torque slightly overestimates its effect compared to calculating the total dust torque based on the entire dust distribution. Nonetheless, in all scenarios, the dust torque contributes to delaying inward planet migration relative to cases where the dust torque is not taken into account.
 
\bibliography{biblio}{}

\begin{thebibliography}{}
\expandafter\ifx\csname natexlab\endcsname\relax\def\natexlab#1{#1}\fi
\providecommand{\url}[1]{\href{#1}{#1}}
\providecommand{\dodoi}[1]{doi:~\href{http://doi.org/#1}{\nolinkurl{#1}}}
\providecommand{\doeprint}[1]{\href{http://ascl.net/#1}{\nolinkurl{http://ascl.net/#1}}}
\providecommand{\doarXiv}[1]{\href{https://arxiv.org/abs/#1}{\nolinkurl{https://arxiv.org/abs/#1}}}

\bibitem[{{Agostini} {et~al.}(2020){Agostini}, {B{\"o}hmer}, {Bosma}, {Clark},
  {Danninger}, {Fruck}, {Gernh{\"a}user}, {G{\"a}rtner}, {Grant}, {Henningsen},
  {Holzapfel}, {Huber}, {Jenkyns}, {Krauss}, {Krings}, {Kopper},
  {Leism{\"u}ller}, {Leys}, {Macoun}, {Meighen-Berger}, {Michel}, {Moore},
  {Morley}, {Padovani}, {Papp}, {Pirenne}, {Qiu}, {Rea}, {Resconi}, {Round},
  {Ruskey}, {Spannfellner}, {Traxler}, {Turcati}, \&
  {Yanez}}]{BorexinoNeutrinos2020}
{Agostini}, M., {B{\"o}hmer}, M., {Bosma}, J., {et~al.} 2020, Nature Astronomy,
  4, 913, \dodoi{10.1038/s41550-020-1182-4}

\bibitem[{{Alibert} {et~al.}(2005){Alibert}, {Mordasini}, {Benz}, \&
  {Winisdoerffer}}]{Alibert05}
{Alibert}, Y., {Mordasini}, C., {Benz}, W., \& {Winisdoerffer}, C. 2005, \aap,
  434, 343, \dodoi{10.1051/0004-6361:20042032}

\bibitem[{{Andrews} {et~al.}(2010){Andrews}, {Wilner}, {Hughes}, {Qi}, \&
  {Dullemond}}]{Andrews10}
{Andrews}, S.~M., {Wilner}, D.~J., {Hughes}, A.~M., {Qi}, C., \& {Dullemond},
  C.~P. 2010, \apj, 723, 1241, \dodoi{10.1088/0004-637X/723/2/1241}

\bibitem[{{Asplund} {et~al.}(2021){Asplund}, {Amarsi}, \&
  {Grevesse}}]{Asplund2021}
{Asplund}, M., {Amarsi}, A.~M., \& {Grevesse}, N. 2021, \aap, 653, A141,
  \dodoi{10.1051/0004-6361/202140445}

\bibitem[{{Ataiee} {et~al.}(2018){Ataiee}, {Baruteau}, {Alibert}, \&
  {Benz}}]{Ataiee18}
{Ataiee}, S., {Baruteau}, C., {Alibert}, Y., \& {Benz}, W. 2018, \aap, 615,
  A110, \dodoi{10.1051/0004-6361/201732026}

\bibitem[{{Baumann} \& {Bitsch}(2020)}]{Baumann2020}
{Baumann}, T., \& {Bitsch}, B. 2020, \aap, 637, A11,
  \dodoi{10.1051/0004-6361/202037579}

\bibitem[{{Ben{\'{\i}}tez-Llambay} {et~al.}(2015){Ben{\'{\i}}tez-Llambay},
  {Masset}, {Koenigsberger}, \& {Szul{\'a}gyi}}]{Benitez-llambay2015}
{Ben{\'{\i}}tez-Llambay}, P., {Masset}, F., {Koenigsberger}, G., \&
  {Szul{\'a}gyi}, J. 2015, \nat, 520, 63, \dodoi{10.1038/nature14277}

\bibitem[{{Ben{\'\i}tez-Llambay} \& {Pessah}(2018)}]{BL2018}
{Ben{\'\i}tez-Llambay}, P., \& {Pessah}, M.~E. 2018, \apjl, 855, L28,
  \dodoi{10.3847/2041-8213/aab2ae}

\bibitem[{{Birnstiel}(2024)}]{Birnstiel2024}
{Birnstiel}, T. 2024, \araa, 62, 157,
  \dodoi{10.1146/annurev-astro-071221-052705}

\bibitem[{{Birnstiel} {et~al.}(2012){Birnstiel}, {Klahr}, \&
  {Ercolano}}]{Birnstiel12}
{Birnstiel}, T., {Klahr}, H., \& {Ercolano}, B. 2012, \aap, 539, A148,
  \dodoi{10.1051/0004-6361/201118136}

\bibitem[{{Bitsch} {et~al.}(2018){Bitsch}, {Morbidelli}, {Johansen}, {Lega},
  {Lambrechts}, \& {Crida}}]{Bitsch18}
{Bitsch}, B., {Morbidelli}, A., {Johansen}, A., {et~al.} 2018, \aap, 612, A30,
  \dodoi{10.1051/0004-6361/201731931}

\bibitem[{{Bitsch} {et~al.}(2019){Bitsch}, {Raymond}, \& {Izidoro}}]{Bitsch19}
{Bitsch}, B., {Raymond}, S.~N., \& {Izidoro}, A. 2019, \aap, 624, A109,
  \dodoi{10.1051/0004-6361/201935007}

\bibitem[{{Brasser} {et~al.}(2017){Brasser}, {Bitsch}, \&
  {Matsumura}}]{Brasser17}
{Brasser}, R., {Bitsch}, B., \& {Matsumura}, S. 2017, \aj, 153, 222,
  \dodoi{10.3847/1538-3881/aa6ba3}

\bibitem[{{Chrenko} \& {Chametla}(2023)}]{ChrenkoChametla2023}
{Chrenko}, O., \& {Chametla}, R.~O. 2023, \mnras, 524, 2705,
  \dodoi{10.1093/mnras/stad2059}

\bibitem[{{Chrenko} {et~al.}(2024){Chrenko}, {Chametla}, {Masset}, {Baruteau},
  \& {Bro{\v{z}}}}]{Chrenko+2024}
{Chrenko}, O., {Chametla}, R.~O., {Masset}, F.~S., {Baruteau}, C., \&
  {Bro{\v{z}}}, M. 2024, \aap, 690, A41, \dodoi{10.1051/0004-6361/202450922}

\bibitem[{{Cornejo} {et~al.}(2023{\natexlab{a}}){Cornejo}, {Masset},
  {Chametla}, \& {Fromenteau}}]{Cornejo+2023}
{Cornejo}, S., {Masset}, F.~S., {Chametla}, R.~O., \& {Fromenteau}, S.
  2023{\natexlab{a}}, \mnras, 522, 678, \dodoi{10.1093/mnras/stad681}

\bibitem[{{Cornejo} {et~al.}(2023{\natexlab{b}}){Cornejo}, {Masset}, \&
  {S{\'a}nchez-Salcedo}}]{Cornejo+2023b}
{Cornejo}, S., {Masset}, F.~S., \& {S{\'a}nchez-Salcedo}, F.~J.
  2023{\natexlab{b}}, \mnras, 523, 936, \dodoi{10.1093/mnras/stad1476}

\bibitem[{{Dittkrist} {et~al.}(2014){Dittkrist}, {Mordasini}, {Klahr},
  {Alibert}, \& {Henning}}]{Dittkrist2014}
{Dittkrist}, K.-M., {Mordasini}, C., {Klahr}, H., {Alibert}, Y., \& {Henning},
  T. 2014, \aap, 567, A121, \dodoi{10.1051/0004-6361/201322506}

\bibitem[{Drazkowska {et~al.}(2016)Drazkowska, {Alibert}, \&
  {Moore}}]{Drazkowska16}
Drazkowska, J., {Alibert}, Y., \& {Moore}, B. 2016, \aap, 594, A105,
  \dodoi{10.1051/0004-6361/201628983}

\bibitem[{{Drazkowska} {et~al.}(2021){Drazkowska}, {Stammler}, \&
  {Birnstiel}}]{Drazkowska21}
{Drazkowska}, J., {Stammler}, S.~M., \& {Birnstiel}, T. 2021, \aap, 647, A15,
  \dodoi{10.1051/0004-6361/202039925}

\bibitem[{{Drazkowska} {et~al.}(2023){Drazkowska}, {Bitsch}, {Lambrechts},
  {Mulders}, {Harsono}, {Vazan}, {Liu}, {Ormel}, {Kretke}, \&
  {Morbidelli}}]{Drazkowska2023PPVII}
{Drazkowska}, J., {Bitsch}, B., {Lambrechts}, M., {et~al.} 2023, in
  Astronomical Society of the Pacific Conference Series, Vol. 534, Protostars
  and Planets VII, ed. S.~{Inutsuka}, Y.~{Aikawa}, T.~{Muto}, K.~{Tomida}, \&
  M.~{Tamura}, 717, \dodoi{10.48550/arXiv.2203.09759}

\bibitem[{{Dullemond} {et~al.}(2018){Dullemond}, {Birnstiel}, {Huang},
  {Kurtovic}, {Andrews}, {Guzm{\'a}n}, {P{\'e}rez}, {Isella}, {Zhu}, {Benisty},
  {Wilner}, {Bai}, {Carpenter}, {Zhang}, \& {Ricci}}]{Dullemond2018}
{Dullemond}, C.~P., {Birnstiel}, T., {Huang}, J., {et~al.} 2018, \apjl, 869,
  L46, \dodoi{10.3847/2041-8213/aaf742}

\bibitem[{{Flaherty} {et~al.}(2020){Flaherty}, {Hughes}, {Simon}, {Qi}, {Bai},
  {Bulatek}, {Andrews}, {Wilner}, \& {K{\'o}sp{\'a}l}}]{Flaherty2020}
{Flaherty}, K., {Hughes}, A.~M., {Simon}, J.~B., {et~al.} 2020, \apj, 895, 109,
  \dodoi{10.3847/1538-4357/ab8cc5}

\bibitem[{{Guilera} {et~al.}(2023){Guilera}, {Benitez-Llambay}, {Miller
  Bertolami}, \& {Pessah}}]{Guilera+2023}
{Guilera}, O.~M., {Benitez-Llambay}, P., {Miller Bertolami}, M.~M., \&
  {Pessah}, M.~E. 2023, \apj, 953, 97, \dodoi{10.3847/1538-4357/acd2cb}

\bibitem[{{Guilera} {et~al.}(2019){Guilera}, {Cuello}, {Montesinos}, {Miller
  Bertolami}, {Ronco}, {Cuadra}, \& {Masset}}]{Guilera2019}
{Guilera}, O.~M., {Cuello}, N., {Montesinos}, M., {et~al.} 2019, \mnras, 486,
  5690, \dodoi{10.1093/mnras/stz1158}

\bibitem[{{Guilera} {et~al.}(2021){Guilera}, {Miller Bertolami}, {Masset},
  {Cuadra}, {Venturini}, \& {Ronco}}]{Guilera2021}
{Guilera}, O.~M., {Miller Bertolami}, M.~M., {Masset}, F., {et~al.} 2021,
  \mnras, 507, 3638, \dodoi{10.1093/mnras/stab2371}

\bibitem[{{Guilera} {et~al.}(2017){Guilera}, {Miller Bertolami}, \&
  {Ronco}}]{Guilera2017b}
{Guilera}, O.~M., {Miller Bertolami}, M.~M., \& {Ronco}, M.~P. 2017, \mnras,
  471, L16, \dodoi{10.1093/mnrasl/slx095}

\bibitem[{{Guilera} {et~al.}(2020){Guilera}, {S{\'a}ndor}, {Ronco},
  {Venturini}, \& {Miller Bertolami}}]{Guilera20}
{Guilera}, O.~M., {S{\'a}ndor}, Z., {Ronco}, M.~P., {Venturini}, J., \& {Miller
  Bertolami}, M.~M. 2020, \aap, 642, A140, \dodoi{10.1051/0004-6361/202038458}

\bibitem[{{Gundlach} \& {Blum}(2015)}]{Gundlach2015}
{Gundlach}, B., \& {Blum}, J. 2015, \apj, 798, 34,
  \dodoi{10.1088/0004-637X/798/1/34}

\bibitem[{{Hou} \& {Yu}(2024)}]{Hou2024}
{Hou}, Q., \& {Yu}, C. 2024, \apj, 972, 152, \dodoi{10.3847/1538-4357/ad6a5c}

\bibitem[{{Hou} \& {Yu}(2025)}]{Hou2025}
---. 2025, \apj, 979, 185, \dodoi{10.3847/1538-4357/ada15a}

\bibitem[{{Ida} \& {Lin}(2004)}]{IdaLin04}
{Ida}, S., \& {Lin}, D.~N.~C. 2004, \apj, 604, 388, \dodoi{10.1086/381724}

\bibitem[{{Ikoma} {et~al.}(2000){Ikoma}, {Nakazawa}, \& {Emori}}]{Ikoma00}
{Ikoma}, M., {Nakazawa}, K., \& {Emori}, H. 2000, \apj, 537, 1013,
  \dodoi{10.1086/309050}

\bibitem[{{Izidoro} {et~al.}(2021){Izidoro}, {Bitsch}, {Raymond}, {Johansen},
  {Morbidelli}, {Lambrechts}, \& {Jacobson}}]{Izidoro2021}
{Izidoro}, A., {Bitsch}, B., {Raymond}, S.~N., {et~al.} 2021, \aap, 650, A152,
  \dodoi{10.1051/0004-6361/201935336}

\bibitem[{{Jim{\'e}nez} \& {Masset}(2017)}]{jm2017}
{Jim{\'e}nez}, M.~A., \& {Masset}, F.~S. 2017, \mnras, 471, 4917,
  \dodoi{10.1093/mnras/stx1946}

\bibitem[{{Johansen} {et~al.}(2021){Johansen}, {Ronnet}, {Bizzarro},
  {Schiller}, {Lambrechts}, {Nordlund}, \& {Lammer}}]{Johansen2021}
{Johansen}, A., {Ronnet}, T., {Bizzarro}, M., {et~al.} 2021, Science Advances,
  7, eabc0444, \dodoi{10.1126/sciadv.abc0444}

\bibitem[{{Kaufmann} {et~al.}(2025){Kaufmann}, {Guilera}, {Alibert}, \& {San
  Sebasti{\'a}n}}]{Kaufmann+2025}
{Kaufmann}, N., {Guilera}, O.~M., {Alibert}, Y., \& {San Sebasti{\'a}n}, I.~L.
  2025, \aap, 696, A65, \dodoi{10.1051/0004-6361/202452428}

\bibitem[{{Lambrechts} {et~al.}(2014){Lambrechts}, {Johansen}, \&
  {Morbidelli}}]{Lambrechts14}
{Lambrechts}, M., {Johansen}, A., \& {Morbidelli}, A. 2014, \aap, 572, A35,
  \dodoi{10.1051/0004-6361/201423814}

\bibitem[{{Lambrechts} {et~al.}(2019){Lambrechts}, {Morbidelli}, {Jacobson},
  {Johansen}, {Bitsch}, {Izidoro}, \& {Raymond}}]{Lambrechts19}
{Lambrechts}, M., {Morbidelli}, A., {Jacobson}, S.~A., {et~al.} 2019, \aap,
  627, A83, \dodoi{10.1051/0004-6361/201834229}

\bibitem[{{Lega} {et~al.}(2014){Lega}, {Crida}, {Bitsch}, \&
  {Morbidelli}}]{Lega2014}
{Lega}, E., {Crida}, A., {Bitsch}, B., \& {Morbidelli}, A. 2014, \mnras, 440,
  683, \dodoi{10.1093/mnras/stu304}

\bibitem[{{Liu} {et~al.}(2019){Liu}, {Ormel}, \& {Johansen}}]{Liu2019}
{Liu}, B., {Ormel}, C.~W., \& {Johansen}, A. 2019, \aap, 624, A114,
  \dodoi{10.1051/0004-6361/201834174}

\bibitem[{Lodders(2020)}]{Lodder2020}
Lodders, K. 2020, Solar Elemental Abundances, in Online Oxford Research
  Encyclopedia of Planetary Science
  (https://doi.org/10.1093/acrefore/9780190647926.013.145). arXiv:1912.00844,
  Oxford University Press, \dodoi{10.1093/acrefore/9780190647926.013.145}

\bibitem[{{Morbidelli} {et~al.}(2015){Morbidelli}, {Lambrechts}, {Jacobson}, \&
  {Bitsch}}]{Morby15}
{Morbidelli}, A., {Lambrechts}, M., {Jacobson}, S., \& {Bitsch}, B. 2015,
  \icarus, 258, 418, \dodoi{10.1016/j.icarus.2015.06.003}

\bibitem[{{Mordasini} \& {Burn}(2024)}]{MordasiniBurn2024}
{Mordasini}, C., \& {Burn}, R. 2024, Reviews in Mineralogy and Geochemistry,
  90, 55, \dodoi{10.2138/rmg.2024.90.03}

\bibitem[{{Musiolik}(2021)}]{Musiolik2021}
{Musiolik}, G. 2021, \mnras, 506, 5153, \dodoi{10.1093/mnras/stab1963}

\bibitem[{{Musiolik} \& {Wurm}(2019)}]{MusiolikWurm2019}
{Musiolik}, G., \& {Wurm}, G. 2019, \apj, 873, 58,
  \dodoi{10.3847/1538-4357/ab0428}

\bibitem[{{Ogihara} \& {Hori}(2020)}]{Ogihara2020}
{Ogihara}, M., \& {Hori}, Y. 2020, \apj, 892, 124,
  \dodoi{10.3847/1538-4357/ab7fa7}

\bibitem[{{Ormel}(2024)}]{Ormel2024}
{Ormel}, C.~W. 2024, arXiv e-prints, arXiv:2411.14643,
  \dodoi{10.48550/arXiv.2411.14643}

\bibitem[{{Paardekooper} {et~al.}(2011){Paardekooper}, {Baruteau}, \&
  {Kley}}]{Paardekooper2011}
{Paardekooper}, S.-J., {Baruteau}, C., \& {Kley}, W. 2011, \mnras, 410, 293,
  \dodoi{10.1111/j.1365-2966.2010.17442.x}

\bibitem[{{Reg{\'a}ly}(2020)}]{Regaly2020}
{Reg{\'a}ly}, Z. 2020, \mnras, 497, 5540, \dodoi{10.1093/mnras/staa2181}

\bibitem[{{Reg{\'a}ly} {et~al.}(2025){Reg{\'a}ly}, {N{\'e}meth},
  {Krup{\'a}nszky}, \& {S{\'a}ndor}}]{Regaly2025}
{Reg{\'a}ly}, Z., {N{\'e}meth}, A., {Krup{\'a}nszky}, G., \& {S{\'a}ndor}, Z.
  2025, \aap, 694, A279, \dodoi{10.1051/0004-6361/202452806}

\bibitem[{{Ronco} {et~al.}(2017){Ronco}, {Guilera}, \& {de
  El{\'\i}a}}]{Ronco2017}
{Ronco}, M.~P., {Guilera}, O.~M., \& {de El{\'\i}a}, G.~C. 2017, \mnras, 471,
  2753, \dodoi{10.1093/mnras/stx1746}

\bibitem[{{Tanigawa} \& {Ikoma}(2007)}]{Tanigawa07}
{Tanigawa}, T., \& {Ikoma}, M. 2007, \apj, 667, 557, \dodoi{10.1086/520499}

\bibitem[{{Toomre}(1964)}]{Toomre1964}
{Toomre}, A. 1964, \apj, 139, 1217, \dodoi{10.1086/147861}

\bibitem[{{Venturini} {et~al.}(2020{\natexlab{a}}){Venturini}, {Guilera},
  {Haldemann}, {Ronco}, \& {Mordasini}}]{Venturini20Letter}
{Venturini}, J., {Guilera}, O.~M., {Haldemann}, J., {Ronco}, M.~P., \&
  {Mordasini}, C. 2020{\natexlab{a}}, \aap, 643, L1,
  \dodoi{10.1051/0004-6361/202039141}

\bibitem[{{Venturini} {et~al.}(2020{\natexlab{b}}){Venturini}, {Guilera},
  {Ronco}, \& {Mordasini}}]{Venturini20ST}
{Venturini}, J., {Guilera}, O.~M., {Ronco}, M.~P., \& {Mordasini}, C.
  2020{\natexlab{b}}, \aap, 644, A174, \dodoi{10.1051/0004-6361/202039140}

\bibitem[{{Venturini} {et~al.}(2020{\natexlab{c}}){Venturini}, {Ronco}, \&
  {Guilera}}]{Venturini20Review}
{Venturini}, J., {Ronco}, M.~P., \& {Guilera}, O.~M. 2020{\natexlab{c}}, \ssr,
  216, 86, \dodoi{10.1007/s11214-020-00700-y}

\bibitem[{{Voelkel} {et~al.}(2022){Voelkel}, {Klahr}, {Mordasini}, \&
  {Emsenhuber}}]{Voelkel2022}
{Voelkel}, O., {Klahr}, H., {Mordasini}, C., \& {Emsenhuber}, A. 2022, \aap,
  666, A90, \dodoi{10.1051/0004-6361/202141830}

\bibitem[{{von Steiger} \& {Zurbuchen}(2016)}]{vonS2016}
{von Steiger}, R., \& {Zurbuchen}, T.~H. 2016, \apj, 816, 13,
  \dodoi{10.3847/0004-637X/816/1/13}

\end{thebibliography}
\bibliographystyle{aasjournal}

\end{document}